\documentclass[runningheads]{llncs}

\usepackage{eccv}

\usepackage{eccvabbrv}

\usepackage{graphicx}
\usepackage{booktabs}

\usepackage[accsupp]{axessibility}  

\usepackage[pagebackref,breaklinks,colorlinks,citecolor=eccvblue]{hyperref}

\usepackage{orcidlink}

\begin{document}

\title{Image and Video Compression using Generative Sparse Representation with Fidelity Controls} 

\titlerunning{Abbreviated paper title}

\author{Wei Jiang\and
Wei Wang}

\authorrunning{W.~Jiang et al.}

\institute{Futurewei Technologies Inc., Santa Clara CA 95050, USA 
\email{\{wjiang,rickweiwang\}@futurewei.com}}

\maketitle

\begin{abstract}

We propose a framework for learned image and video compression using the generative sparse visual representation (SVR) guided by fidelity-preserving controls. By embedding inputs into  a discrete latent space spanned by learned visual codebooks, SVR-based compression transmits integer codeword indices, which is efficient and cross-platform robust.  However, high-quality (HQ) reconstruction in the decoder relies on intermediate feature inputs from the encoder via direct connections. Due to the  prohibitively high transmission costs, previous SVR-based compression methods remove such feature links, resulting in largely degraded reconstruction quality. In this work, we treat the intermediate features as fidelity-preserving control signals that guide the conditioned generative reconstruction in the decoder. Instead of discarding or directly transferring such signals, we draw them from a low-quality (LQ) fidelity-preserving alternative input that is sent to the decoder with very low bitrate. These control signals provide complementary fidelity cues to improve reconstruction, and their quality is determined by the compression rate of the LQ alternative, which can be tuned to trade off bitrate, fidelity and perceptual quality. Our framework can be conveniently used for both learned image compression (LIC) and learned video compression (LVC). Since SVR is robust against input perturbations, a large portion of codeword indices between adjacent frames can be the same. By only transferring different indices, SVR-based LIC and LVC can share a similar processing pipeline.  Experiments over standard image and video compression benchmarks demonstrate the effectiveness of our approach.

\end{abstract}

\section{Introduction}

Image and video compression has been a decades-long research topic, and great success has been achieved recently by using neural networks (NN) for both learned image compression (LIC) \cite{Cheng2020,MLIC2023} and learned video compression (LVC)  \cite{FVCCVPR2021,AlphaVC2022}. 
Most existing LIC methods \cite{Cheng2020,JPEGAI,MLIC2023} use a hyperprior framework  \cite{hyperprior}, which combines classical entropy coding with NN-based representation learning in a Variational AutoEncoder (VAE) structure. An entropy model is used to encode the quantized latent feature for easy transmission. Most existing LVC methods follow the pipeline of traditional video coding \cite{hevc_std,vvc_std}, while replacing processing modules like motion estimation, motion compensation, residue coding \textit{etc.} by learned NNs. 

In this paper, we explore a different compression pipeline for both LIC and LVC, based on controlled generative modeling using the Sparse Visual Representation (SVR) (as shown in Fig.~\ref{fig:svr_general} and Fig.~\ref{fig:svr-lq-control}). We learn discrete generative priors as visual codebooks, and embed images into a discrete latent space spanned by the codebooks. By sharing the learned codebooks between the encoder and decoder, images can be mapped to integer codeword indices in the encoder, and the decoder can use these indices to retrieve the corresponding codewords' latent features for reconstruction. 

The SVR-based compression has several benefits. (1) Transferring integer indices is very robust to heterogeneous platforms. One caveat of the hyperprior framework is the extreme sensitivity to small differences between the encoder and decoder in calculating the hyperpriors \cite{entropyerrorICLR2019}. Even perturbations caused by floating round-off error can lead to catastrophic error propagation in the decoded latent feature. By encoding codeword indices instead of latent features, SVR-based compression does not suffer from such sensitivity. (2) Transferring indices gives the freedom of expanding latent feature dimension (often associated with better representation power for better reconstruction) without increasing bitrate, in comparison to transferring latent features or residues. (3) Generative SVR increases robustness to input degradations. Realistic and rich textures can be generated using hiqh-quality (HQ) codebooks even for low-quality (LQ) inputs. 

\begin{figure*}[t]

    \centering
    \begin{minipage}[b][0.13\linewidth][b]{0.2\linewidth}
    \centering
    \centerline{\scriptsize{\textcolor{white}{g}$3680\!\times\!2456$\textcolor{white}{g}}}
    \vspace{0.03cm}
    \includegraphics[width=\textwidth,height=0.645\linewidth]{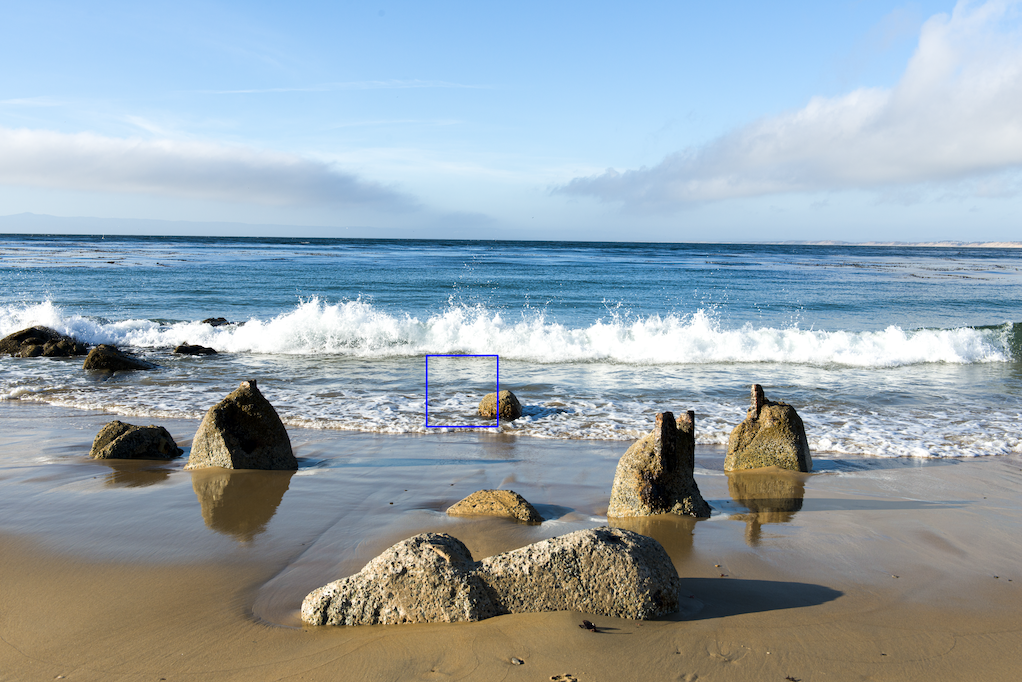}
    \vspace{-0.3cm}
    \centerline{\tiny\textcolor{white}{0.413$\mid$}}\medskip
    \end{minipage}
    \hspace{0.15cm}
    \begin{minipage}[b]{0.13\linewidth}
    \centering
    \centerline{\scriptsize \textcolor{white}{g} \textcolor{white}{g}}
    \vspace{0.02cm}
    \includegraphics[width=\textwidth]{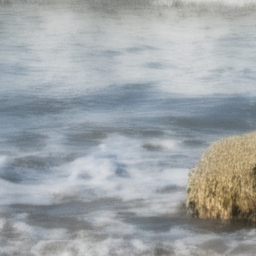}
    \vspace{-0.3cm}
    \centerline{\tiny 0.282$\mid$24.4$\mid$0.782}\medskip
    \end{minipage}
    \hspace{0.15cm}
    \begin{minipage}[b]{0.13\linewidth}
    \centering
    \centerline{\scriptsize \textcolor{white}{g} \textcolor{white}{g}}
    \vspace{0.02cm}
    \includegraphics[width=\textwidth]{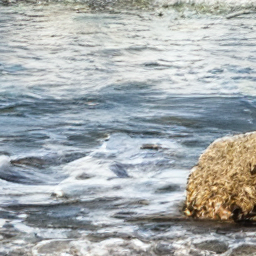}
    \vspace{-0.3cm}
    \centerline{\tiny 0.115$\mid$24.6$\mid$0.722}\medskip
    \end{minipage}
    \hspace{0.15cm}
    \begin{minipage}[b]{0.13\linewidth}
    \centering
    \centerline{\scriptsize  \textcolor{white}{g} \textcolor{white}{g}}
    \vspace{0.02cm}
    \includegraphics[width=\textwidth]{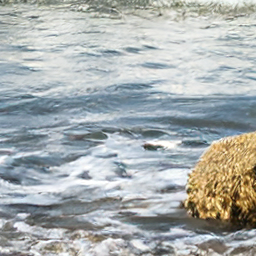}
    \vspace{-0.3cm}
    \centerline{\tiny 0.149$\mid${27.0}$\mid$0.816}\medskip
    \end{minipage}
    \hspace{0.15cm}
    \begin{minipage}[b]{0.13\linewidth}
    \centering
    \centerline{\scriptsize \textcolor{white}{g} \textcolor{white}{g}}
    \vspace{0.02cm}
    \includegraphics[width=\textwidth]{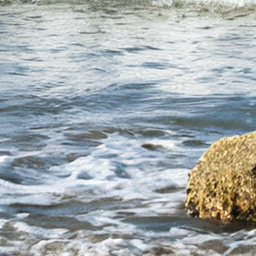}
    \vspace{-0.3cm}
    \centerline{\tiny {0.105}$\mid$30.4$\mid$0.861}\medskip
    \end{minipage}
    \hspace{0.15cm}
    \begin{minipage}[b]{0.13\linewidth}
    \centering
    \centerline{\scriptsize \textcolor{white}{g} \textcolor{white}{g}}
    \vspace{0.02cm}
    \includegraphics[width=\textwidth]{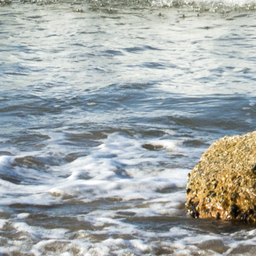}
    \vspace{-0.3cm}
    \centerline{\tiny\textcolor{white}{0.413$\mid$}}\medskip
    \end{minipage}    
    \vspace{.1em}

    \centering
    \begin{minipage}[b][0.13\linewidth][b]{0.2\linewidth}
    \centering
    \centerline{\scriptsize{\textcolor{white}{g}$2096\!\times\!1400$\textcolor{white}{g}}}
    \vspace{0.02cm}
    \includegraphics[width=\textwidth,height=0.645\linewidth]{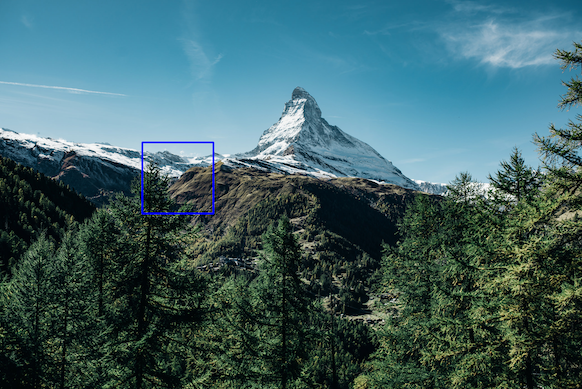}
    \vspace{-0.3cm}
    \centerline{\tiny\textcolor{white}{0.413$\mid$}}\medskip
    \end{minipage}
    \hspace{0.15cm}
    \begin{minipage}[b]{0.13\linewidth}
    \centering
    \centerline{\scriptsize \textcolor{white}{g} \textcolor{white}{g}}
    \vspace{0.02cm}
    \includegraphics[width=\textwidth]{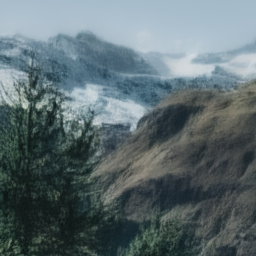}
    \vspace{-0.3cm}
    \centerline{\tiny 0.305$\mid$21.3$\mid$0.662}\medskip
    \end{minipage}
    \hspace{0.15cm}
    \begin{minipage}[b]{0.13\linewidth}
    \centering
    \centerline{\scriptsize \textcolor{white}{g} \textcolor{white}{g}}
    \vspace{0.02cm}
    \includegraphics[width=\textwidth]{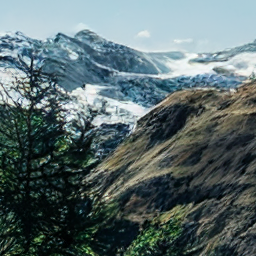}
    \vspace{-0.3cm}
    \centerline{\tiny 0.155$\mid$19.8$\mid$0.620}\medskip
    \end{minipage}
    \hspace{0.15cm}
    \begin{minipage}[b]{0.13\linewidth}
    \centering
    \centerline{\scriptsize  \textcolor{white}{g}\textcolor{white}{g}}
    \vspace{0.02cm}
    \includegraphics[width=\textwidth]{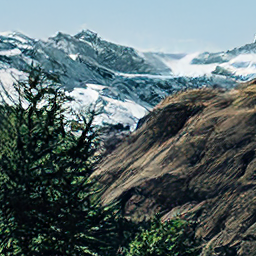}
    \vspace{-0.3cm}
    \centerline{\tiny 0.144$\mid${22.0}$\mid$0.753}\medskip
    \end{minipage}
    \hspace{0.15cm}
    \begin{minipage}[b]{0.13\linewidth}
    \centering
    \centerline{\scriptsize \textcolor{white}{g} \textcolor{white}{g}}
    \vspace{0.02cm}
    \includegraphics[width=\textwidth]{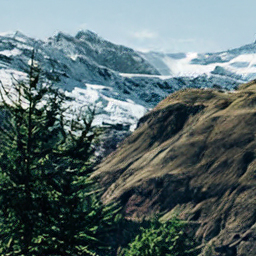}
    \vspace{-0.3cm}
    \centerline{\tiny {0.083}$\mid$26.9$\mid$0.874}\medskip
    \end{minipage}
    \hspace{0.15cm}
    \begin{minipage}[b]{0.13\linewidth}
    \centering
    \centerline{\scriptsize \textcolor{white}{g} \textcolor{white}{g}}
    \vspace{0.02cm}
    \includegraphics[width=\textwidth]{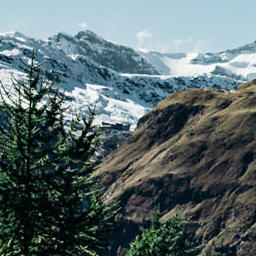}
    \vspace{-0.3cm}
    \centerline{\tiny\textcolor{white}{0.413$\mid$}}\medskip
    \end{minipage}    
    \vspace{.1em}
    
   \centering
    \begin{minipage}[b][0.13\linewidth][b]{0.2\linewidth}
    \centering
    \centerline{\scriptsize{\textcolor{white}{g}$1920\!\times\!1080$\textcolor{white}{g}}}
    \vspace{0.02cm}
    \includegraphics[width=\textwidth,height=0.645\linewidth]{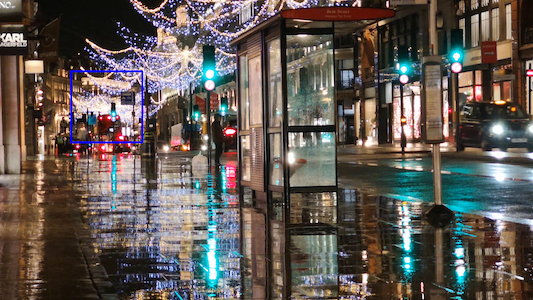}
    \vspace{-0.3cm}
    \centerline{\tiny\textcolor{white}{0.413$\mid$}}\medskip
    \end{minipage}
    \hspace{0.15cm}
    \begin{minipage}[b]{0.13\linewidth}
    \centering
    \centerline{\scriptsize \textcolor{white}{g} \textcolor{white}{g}}
    \vspace{0.02cm}
    \includegraphics[width=\textwidth]{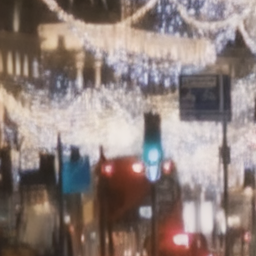}
    \vspace{-0.3cm}
    \centerline{\tiny 0.331$\mid$22.5$\mid$0.745}\medskip
    \end{minipage}
    \hspace{0.15cm}
    \begin{minipage}[b]{0.13\linewidth}
    \centering
    \centerline{\scriptsize \textcolor{white}{g} \textcolor{white}{g}}
    \vspace{0.02cm}
    \includegraphics[width=\textwidth]{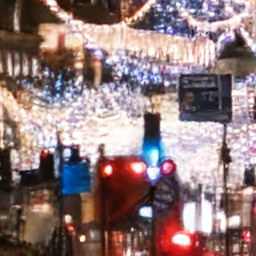}
    \vspace{-0.3cm}
    \centerline{\tiny 0.160$\mid$23.2$\mid$0.742}\medskip
    \end{minipage}
    \hspace{0.15cm}
    \begin{minipage}[b]{0.13\linewidth}
    \centering
    \centerline{\scriptsize  \textcolor{white}{g} \textcolor{white}{g}}
    \vspace{0.02cm}
    \includegraphics[width=\textwidth]{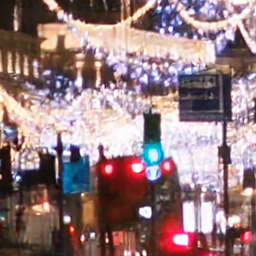}
    \vspace{-0.3cm}
    \centerline{\tiny 0.142$\mid${26.6}$\mid$0.838}\medskip
    \end{minipage}
    \hspace{0.15cm}
    \begin{minipage}[b]{0.13\linewidth}
    \centering
    \centerline{\scriptsize \textcolor{white}{g} \textcolor{white}{g}}
    \vspace{0.02cm}
    \includegraphics[width=\textwidth]{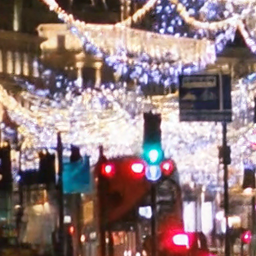}
    \vspace{-0.3cm}
    \centerline{\tiny {0.083}$\mid$30.3$\mid$0.903}\medskip
    \end{minipage}
    \hspace{0.15cm}
    \begin{minipage}[b]{0.13\linewidth}
    \centering
    \centerline{\scriptsize \textcolor{white}{g} \textcolor{white}{g}}
    \vspace{0.02cm}
    \includegraphics[width=\textwidth]{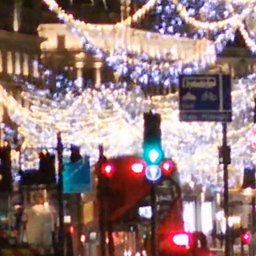}
    \vspace{-0.3cm}
    \centerline{\tiny\textcolor{white}{0.413$\mid$}}\medskip
    \end{minipage}    
    \vspace{.1em}

   \centering
    \begin{minipage}[b][0.13\linewidth][b]{0.2\linewidth}
    \centering
    \centerline{\scriptsize{\textcolor{white}{g}$2144\!\times\!1424$\textcolor{white}{g}}}
    \vspace{0.02cm}
    \includegraphics[width=\textwidth,height=0.645\linewidth]{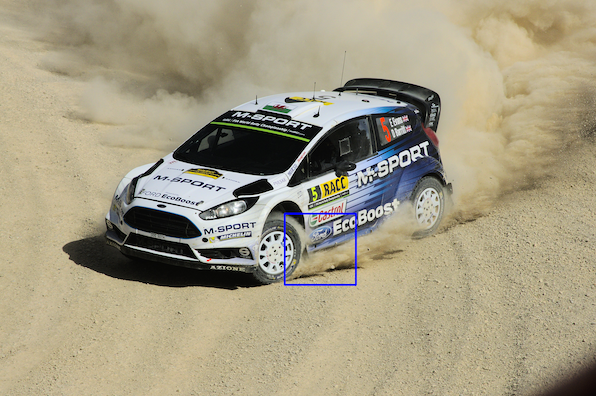}
    \vspace{-0.3cm}
    \centerline{\tiny\textcolor{white}{0.413$\mid$}}\medskip
    \end{minipage}
    \hspace{0.15cm}
    \begin{minipage}[b]{0.13\linewidth}
    \centering
    \centerline{\scriptsize \textcolor{white}{g} \textcolor{white}{g}}
    \vspace{0.02cm}
    \includegraphics[width=\textwidth]{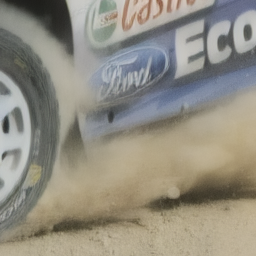}
    \vspace{-0.3cm}
    \centerline{\tiny 0.372$\mid$23.3$\mid$0.671}\medskip
    \end{minipage}
    \hspace{0.15cm}
    \begin{minipage}[b]{0.13\linewidth}
    \centering
    \centerline{\scriptsize \textcolor{white}{g} \textcolor{white}{g}}
    \vspace{0.02cm}
    \includegraphics[width=\textwidth]{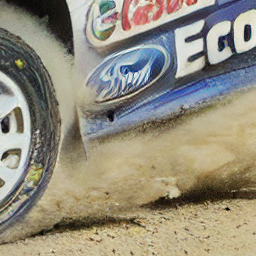}
    \vspace{-0.3cm}
    \centerline{\tiny 0.153$\mid$23.3$\mid$0.594}\medskip
    \end{minipage}
    \hspace{0.15cm}
    \begin{minipage}[b]{0.13\linewidth}
    \centering
    \centerline{\scriptsize  \textcolor{white}{g} \textcolor{white}{g}}
    \vspace{0.02cm}
    \includegraphics[width=\textwidth]{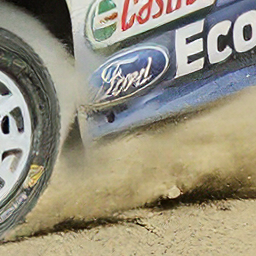}
    \vspace{-0.3cm}
    \centerline{\tiny 0.208$\mid${24.9}$\mid$0.685}\medskip
    \end{minipage}
    \hspace{0.15cm}
    \begin{minipage}[b]{0.13\linewidth}
    \centering
    \centerline{\scriptsize \textcolor{white}{g} \textcolor{white}{g}}
    \vspace{0.02cm}
    \includegraphics[width=\textwidth]{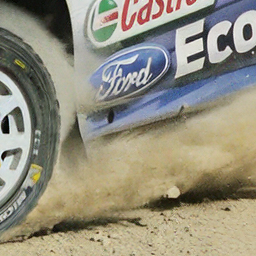}
    \vspace{-0.3cm}
    \centerline{\tiny {0.141}$\mid$28.4$\mid$0.748}\medskip
    \end{minipage}
    \hspace{0.15cm}
    \begin{minipage}[b]{0.13\linewidth}
    \centering
    \centerline{\scriptsize \textcolor{white}{g} \textcolor{white}{g}}
    \vspace{0.02cm}
    \includegraphics[width=\textwidth]{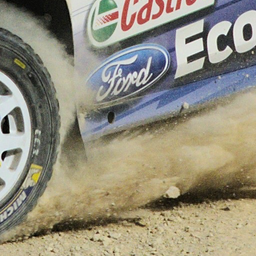}
    \vspace{-0.3cm}
    \centerline{\tiny\textcolor{white}{0.413$\mid$}}\medskip
    \end{minipage}    
    \vspace{.1em}

   \centering
    \begin{minipage}[b][0.13\linewidth][b]{0.2\linewidth}
    \centering
    \centerline{\scriptsize{\textcolor{white}{g}$1984\!\times\!1320$\textcolor{white}{g}}}
    \vspace{0.02cm}
    \includegraphics[width=\textwidth,height=0.645\linewidth]{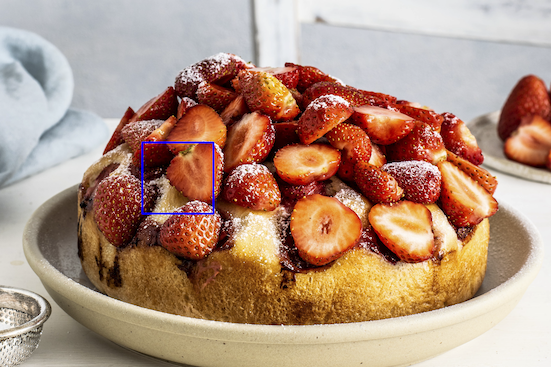}
    \vspace{-0.3cm}
    \centerline{\tiny\textcolor{white}{0.413$\mid$}}\medskip
    \end{minipage}
    \hspace{0.15cm}
    \begin{minipage}[b]{0.13\linewidth}
    \centering
    \centerline{\scriptsize \textcolor{white}{g} \textcolor{white}{g}}
    \vspace{0.02cm}
    \includegraphics[width=\textwidth]{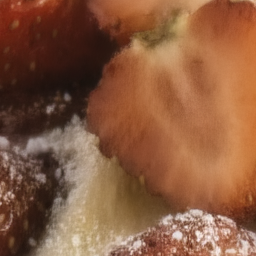}
    \vspace{-0.3cm}
    \centerline{\tiny 0.279$\mid$23.6$\mid$0.796}\medskip
    \end{minipage}
    \hspace{0.15cm}
    \begin{minipage}[b]{0.13\linewidth}
    \centering
    \centerline{\scriptsize \textcolor{white}{g} \textcolor{white}{g}}
    \vspace{0.02cm}
    \includegraphics[width=\textwidth]{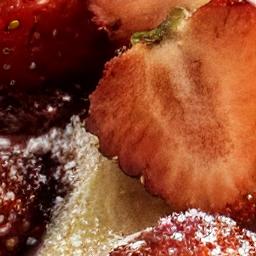}
    \vspace{-0.3cm}
    \centerline{\tiny 0.120$\mid$23.7$\mid$0.719}\medskip
    \end{minipage}
    \hspace{0.15cm}
    \begin{minipage}[b]{0.13\linewidth}
    \centering
    \centerline{\scriptsize  \textcolor{white}{g} \textcolor{white}{g}}
    \vspace{0.02cm}
    \includegraphics[width=\textwidth]{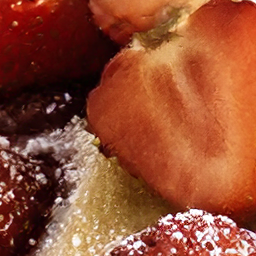}
    \vspace{-0.3cm}
    \centerline{\tiny 0.143$\mid${26.6}$\mid$0.799}\medskip
    \end{minipage}
    \hspace{0.15cm}
    \begin{minipage}[b]{0.13\linewidth}
    \centering
    \centerline{\scriptsize \textcolor{white}{g} \textcolor{white}{g}}
    \vspace{0.02cm}
    \includegraphics[width=\textwidth]{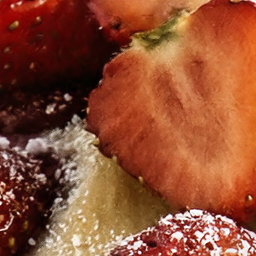}
    \vspace{-0.3cm}
    \centerline{\tiny {0.088}$\mid$30.2$\mid$0.856}\medskip
    \end{minipage}
    \hspace{0.15cm}
    \begin{minipage}[b]{0.13\linewidth}
    \centering
    \centerline{\scriptsize \textcolor{white}{g} \textcolor{white}{g}}
    \vspace{0.02cm}
    \includegraphics[width=\textwidth]{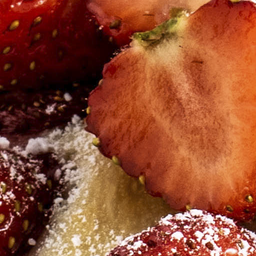}
    \vspace{-0.3cm}
    \centerline{\tiny\textcolor{white}{0.413$\mid$}}\medskip
    \end{minipage}    
    \vspace{.1em}

   \centering
    \begin{minipage}[b][0.13\linewidth][b]{0.2\linewidth}
    \centering
    \centerline{\scriptsize{\textcolor{white}{g}$1472\!\times\!976$\textcolor{white}{g}}}
    \vspace{0.02cm}
    \includegraphics[width=\textwidth,height=0.645\linewidth]{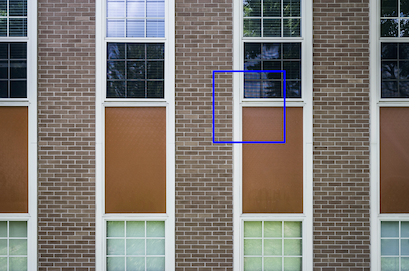}
    \vspace{-0.3cm}
    \centerline{\tiny\textcolor{white}{0.413$\mid$}}\medskip
    \end{minipage}
    \hspace{0.15cm}
    \begin{minipage}[b]{0.13\linewidth}
    \centering
    \centerline{\scriptsize \textcolor{white}{g} \textcolor{white}{g}}
    \vspace{0.02cm}
    \includegraphics[width=\textwidth]{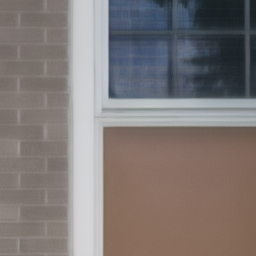}
    \vspace{-0.3cm}
    \centerline{\tiny 0.356$\mid$21.4$\mid$0.631}\medskip
    \end{minipage}
    \hspace{0.15cm}
    \begin{minipage}[b]{0.13\linewidth}
    \centering
    \centerline{\scriptsize \textcolor{white}{g} \textcolor{white}{g}}
    \vspace{0.02cm}
    \includegraphics[width=\textwidth]{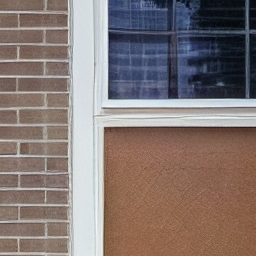}
    \vspace{-0.3cm}
    \centerline{\tiny 0.099$\mid$24.8$\mid$0.739}\medskip
    \end{minipage}
    \hspace{0.15cm}
    \begin{minipage}[b]{0.13\linewidth}
    \centering
    \centerline{\scriptsize  \textcolor{white}{g} \textcolor{white}{g}}
    \vspace{0.02cm}
    \includegraphics[width=\textwidth]{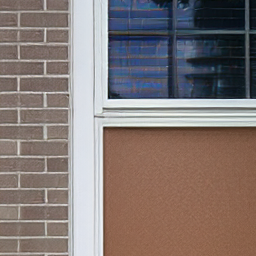}
    \vspace{-0.3cm}
    \centerline{\tiny 0.089$\mid${27.5}$\mid$0.831}\medskip
    \end{minipage}
    \hspace{0.15cm}
    \begin{minipage}[b]{0.13\linewidth}
    \centering
    \centerline{\scriptsize \textcolor{white}{g} \textcolor{white}{g}}
    \vspace{0.02cm}
    \includegraphics[width=\textwidth]{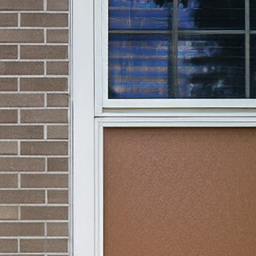}
    \vspace{-0.3cm}
    \centerline{\tiny {0.049}$\mid$31.2$\mid$0.878}\medskip
    \end{minipage}
    \hspace{0.15cm}
    \begin{minipage}[b]{0.13\linewidth}
    \centering
    \centerline{\scriptsize \textcolor{white}{g} \textcolor{white}{g}}
    \vspace{0.02cm}
    \includegraphics[width=\textwidth]{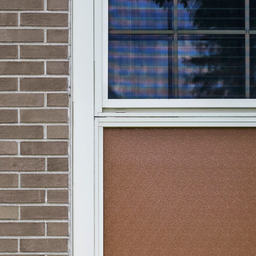}
    \vspace{-0.3cm}
    \centerline{\tiny\textcolor{white}{0.413$\mid$}}\medskip
    \end{minipage}    
    \vspace{.1em}    

   \centering
    \begin{minipage}[b][0.13\linewidth][b]{0.2\linewidth}
    \centering
    \centerline{\scriptsize{\textcolor{white}{g}$1192\!\times\!832$\textcolor{white}{g}}}
    \vspace{0.02cm}
    \includegraphics[width=\textwidth,height=0.645\linewidth]{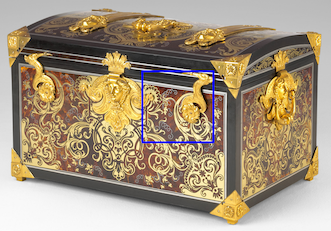}
    \vspace{-0.3cm}
    \centerline{\tiny\textcolor{white}{0.413$\mid$}}\medskip
    \end{minipage}
    \hspace{0.15cm}
    \begin{minipage}[b]{0.13\linewidth}
    \centering
    \centerline{\scriptsize \textcolor{white}{g} \textcolor{white}{g}}
    \vspace{0.02cm}
    \includegraphics[width=\textwidth]{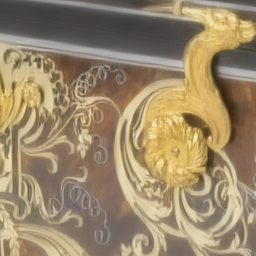}
    \vspace{-0.3cm}
    \centerline{\tiny 0.245$\mid$20.8$\mid$0.719}\medskip
    \end{minipage}
    \hspace{0.15cm}
    \begin{minipage}[b]{0.13\linewidth}
    \centering
    \centerline{\scriptsize \textcolor{white}{g} \textcolor{white}{g}}
    \vspace{0.02cm}
    \includegraphics[width=\textwidth]{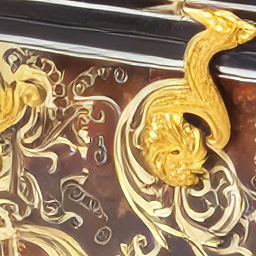}
    \vspace{-0.3cm}
    \centerline{\tiny 0.127$\mid$20.3$\mid$0.675}\medskip
    \end{minipage}
    \hspace{0.15cm}
    \begin{minipage}[b]{0.13\linewidth}
    \centering
    \centerline{\scriptsize  \textcolor{white}{g} \textcolor{white}{g}}
    \vspace{0.02cm}
    \includegraphics[width=\textwidth]{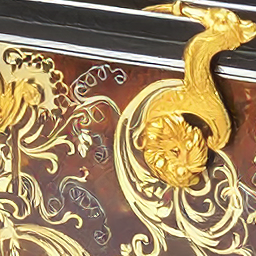}
    \vspace{-0.3cm}
    \centerline{\tiny 0.111$\mid${23.4}$\mid$0.798}\medskip
    \end{minipage}
    \hspace{0.15cm}
    \begin{minipage}[b]{0.13\linewidth}
    \centering
    \centerline{\scriptsize \textcolor{white}{g} \textcolor{white}{g}}
    \vspace{0.02cm}
    \includegraphics[width=\textwidth]{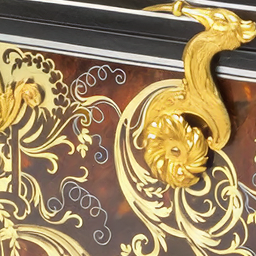}
    \vspace{-0.3cm}
    \centerline{\tiny {0.049}$\mid$28.8$\mid$0.907}\medskip
    \end{minipage}
    \hspace{0.15cm}
    \begin{minipage}[b]{0.13\linewidth}
    \centering
    \centerline{\scriptsize \textcolor{white}{g} \textcolor{white}{g}}
    \vspace{0.02cm}
    \includegraphics[width=\textwidth]{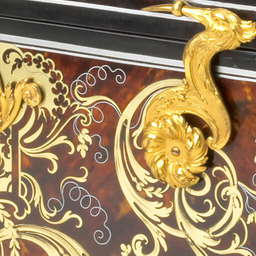}
    \vspace{-0.3cm}
    \centerline{\tiny\textcolor{white}{0.413$\mid$}}\medskip
    \end{minipage}    
\caption{``LPIPS$|$PSNR$|$SSIM'' under each result. ``$\text{SVR}_{\text{m}}$''/ ``$\text{SVR}_{\text{s}}$'': multi-codebook-based/single-codebook-based SVR. ``LQ''/``LQ$\downarrow\!2\!\times$''/``LQ$\downarrow\!4\!\times$'': $x^{LQ}$ of original size/$x^{LQ}$ with $2\!\times$ downsampling-upsampling /$x^{LQ}$ with $4\!\times$ downsampling-upsampling.}
\label{fig:LIC-example}
\end{figure*}

However, SVR-based HQ restoration \cite{AdaCode,GLEAN2021,FeMaSR} relies on the dense connection of multi-scale features between the embedding network (encoder) and reconstruction network (decoder). Such intermediate features are too large to transfer, defeating the purpose of the compression task. As a result, previous SVR-based compression methods remove such direct feature links. By applying to specific content like human faces \cite{NTIREFaceCompress,CodeFormer2022}, a code transformer is used to recover an aligned structured code sequence for HQ face restoration without intermediate features. For general images, M-AdaCode \cite{WACVLIC2024} compensates the performance loss of removed feature links by using data-adaptive weights to combine multiple semantic-class-dependent codebooks and uses weight masking to reduce transmitted weight parameters. However, although the restored images may look okay perceptually, important fidelity details are usually lost. As shown in Fig.~\ref{fig:LIC-example}, without direct feature links images generated by M-AdaCode often lack rich details. 

In our opinion, the generative SVR-based reconstruction aims at high perceptual quality, and the multi-scale intermediate features provide complementary fidelity details to the reconstruction. Such details should NOT be ignored for applications like compression. Therefore, we focus on how to obtain effective and transmission-friendly fidelity information to balance bitrate and quality. 

Our work is inspired by the success of ControlNet \cite{ControlNet2023} where conditioning controls are used to guide image generation. We view the multi-scale intermediate features as fidelity-preserving control signals that guide the conditioned reconstruction in the decoder. As control conditions, such signals do NOT have to come from the original input. Instead, we draw these control signals from an LQ alternative of the original input in decoder. This LQ alternative is computed in decoder based on highly compressed easy-to-transmit fidelity-preserving information, which is generated by fidelity-preserving methods like the previous NN-based or traditional image and video compression methods. 

Based on this idea, we propose a framework (Fig.~\ref{fig:svr-lq-control}) that combines generative SVR-based restoration with fidelity-preserving compression. A highly compressed LQ alternative is transmitted with efficient bits, from which LQ control conditions are extracted to guide the reconstruction process. A conditioned generation network with weighted feature modulation is used to combine the SVR-based latent features with the LQ control features. The quality of the LQ control features is determined by the bitrate of the LQ alternative. The strength of the LQ control features in the conditioned generation process balances the importance between the HQ codebook and LQ fidelity details, which can be tuned based on the current reconstruction target to pursue high perceptual quality or high fidelity. As shown in Fig.~\ref{fig:LIC-example}, with our LQ control features, the restored results have largely improved fidelity with rich details.

In addition, we extend the SVR-based LIC framework into an effective LVC framework. Since SVR is robust against input degradation and small perturbations, a substantial amount of codeword indices between adjacent frames can be the same (47\% in our experiments). We only need to transfer different indices for most frames. No motion estimation or motion compensation is involved and there is no error propagation. Comparing to previous LIC and LVC, our SVR-based LIC and LVC share a similar processing pipeline, which makes it possible to simplify industrial productive optimization.

We evaluate our approach using benchmark datasets for image and video standardization. Specifically, SVR-based LIC is tested over the JPEG-AI dataset \cite{JPEG-AI-data}. SVR-base LVC is tested over a combined dataset comprising of video sequences from AOM \cite{aom_std}, MPEG \cite{mpeg_std}, JVET \cite{Jvet_std}, and AVS \cite{avs_std}. Also, we evaluate the performance of different SVR-based restoration methods, based on a single codebook \cite{FeMaSR} or multiple codebooks \cite{AdaCode}. Experimental results demonstrate the effectiveness of our method. 

\section{Related Works}

\subsection{Sparse Visual Representation Learning}
Discrete generative priors have shown impressive performance in image restoration tasks like  super-resolution \cite{FeMaSR}, denoising \cite{VQGAN} \textit{etc}. By embedding images into a discrete latent space spanned by learned visual codebooks, SVR improves robustness to various degradations. For instance, VQ-VAE \cite{VQVAE} learns a highly compressed codebook by a vector-quantized VAE. VQGAN \cite{VQGAN} further improves restoration quality by using GAN with adversarial and perceptual loss. In general, natural images have very complicated content, and it is difficult to learn a single class-agnostic codebook for all image categories. Therefore, most methods focus on specific categories. In particular, great success has been achieved in face generation due to the highly structured characteristics of human faces \cite{RestoreFormer,CodeFormer2022}. 

For general images, to improve the restoration power, the recent AdaCode method \cite{AdaCode} uses an image-adaptive codebook learning approach. Instead of learning a single codebook for all categories of images, a set of basis codebooks are learned, each corresponding to a semantic partition of the latent space. A weight map to combine such basis codebooks is adaptively determined for each input image. By learning the semantic-class-guided codebooks, the semantic-class-agnostic restoration performance can be largely improved. \vspace{-1em}

\subsection{Learned Image Compression} \vspace{-.3em}

There are two main research topics for LIC: how to learn a latent representation, and how to quantize and encode the latent representation. One most popular framework is based on hyperpriors \cite{hyperprior}, where the image is transformed into a dense latent feature, and an entropy model encodes/decodes the quantized latent feature for efficient transmission. Many improvements have been made to improve the transformation for computing the latent \cite{Cheng2020,VCT2022,windowtransformer} and/or the entropy model \cite{checkerboard,VCT2022,hypertransformer}.

One vital issue of the hyperprior framework is the extreme sensitivity to small differences between the encoder and decoder in calculating the hyperpriors \cite{entropyerrorICLR2019}. Even floating round-off error can lead to catastrophic error propagation in the decoded latent feature. Most works simply assume homogeneous platforms and deterministic CPU calculation. Some work uses integer NN to prevent non-deterministic GPU computation \cite{entropyerrorICLR2019}. Some work designs special NN module that is computational friendly to CPU to speed up inference \cite{cpufriendly}. However, such solutions cannot be easily generalized to arbitrary network architectures. 

Also, it is well known that there are complex relations among bitrate, distortion, and perceptual quality \cite{distortionperception,ratedistortionperception}, and it is difficult to pursue high perceptual quality and high pixel-level fidelity at the same time. How to flexibly trade off these factors remains an open issue. \vspace{-1em}

\subsection{Learned Video Compression}\vspace{-.3em}

Existing LVC methods \cite{FVCCVPR2021,DVC2019,LVCICCV2019,AlphaVC2022} follow the traditional video coding pipeline by replacing  processing modules like motion estimation, motion compensation, post-enhancement by NNs. Generally the independent (I) frames in a GoP (group of pictures) are compressed as images, and the predictive (P) frames and the bidirectional predictive (B) frames are compressed based on motion estimation and residue coding. This pipeline is not designed for LVC, resulting in error accumulation from different modules. Also the computation cost is generally very high due to the complicated framework. \vspace{-1em}

\subsection{SVR-based Compression}\vspace{-.3em}

SVR is intuitively suitable for compression, since the integer codeword indices are easy to transfer and are robust to small computation differences in heterogeneous hardware and software platforms. However, HQ SVR-based restoration relies on direct links of multi-scale features between the encoder and decoder. Such features are too expensive to transfer, which often cost more bits than the original input. To make SVR-based compression feasible, previous approaches remove such feature connections. For example, when applied to specific categories like aligned human faces \cite{NTIREFaceCompress,Oneshot2021,CodeFormer2022}, it is possible to predict a cohesive code sequence for HQ restoration without direct feature links. However, for general images the reconstruction quality is severely impacted without such features. As a result, most methods focus on very low-bitrate scenarios \cite{MIMLIC}, where reconstruction with low fidelity yet good perceptual quality is tolerated. The recent M-AdaCode \cite{AdaCode} compensates the performance loss of the removed feature links
by using data-adaptive weights to combine multiple semantic-class-dependent
codebooks and trades off bitrate and distortion by weight masking to reduce transmitted weight parameters. Unfortunately, for general image content, without the feature connections the restoration quality is overall unsatisfactory.

\section{SVR-based Compression with Conditional Controls}

\begin{figure*}
  \centering
   \includegraphics[width=\linewidth]{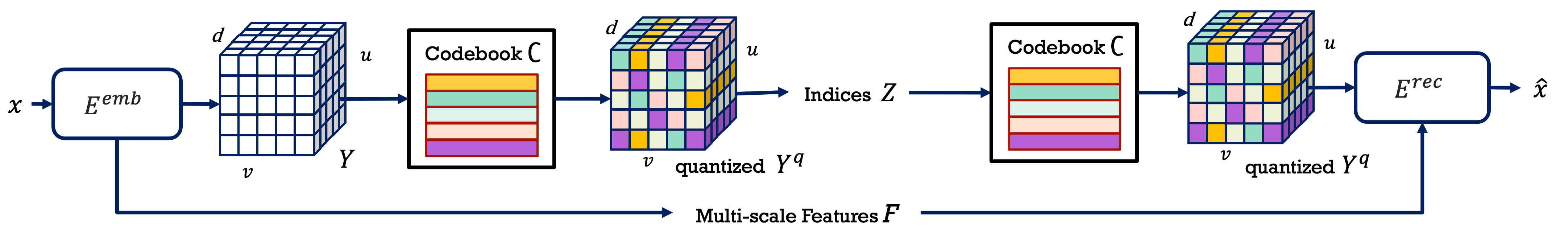}    
   
   \centering{(a) using a single codebook}\vspace{-.5em}
   
   \includegraphics[width=\linewidth]{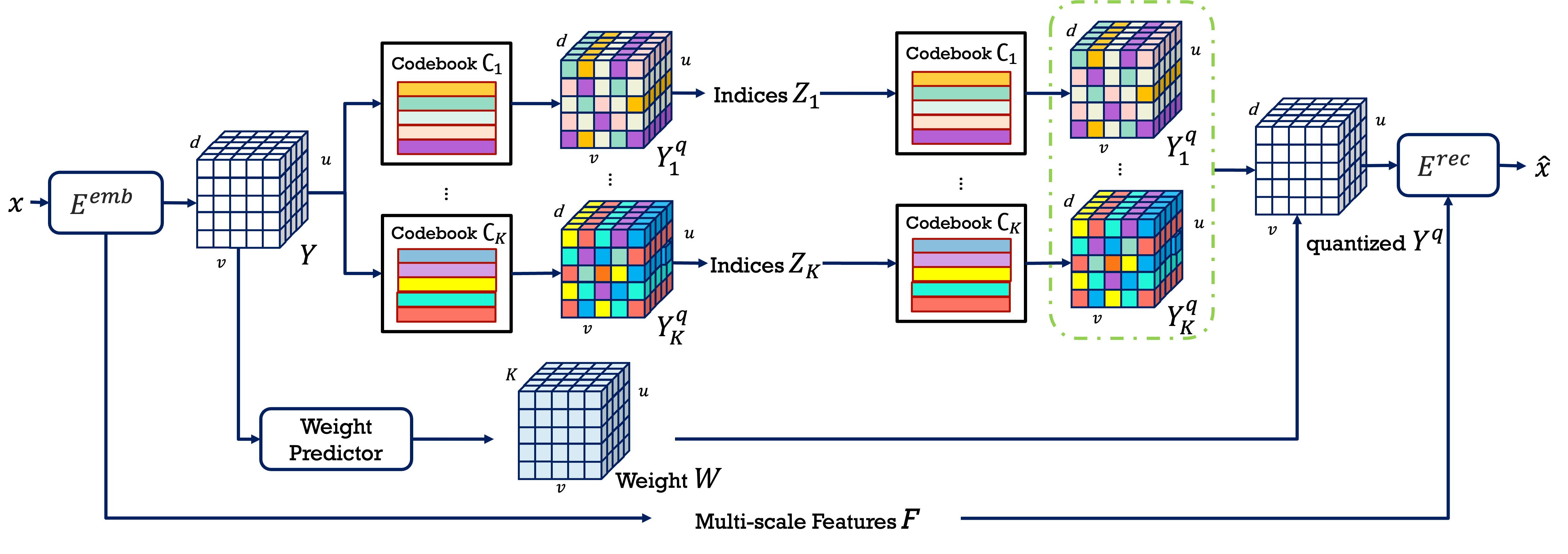}      
   
   \centering{(b) using multiple basis codebooks} \vspace{-.5em}
   \caption{The general workflow of SVR-based restoration. Discarding the multi-scale features $F$ will sacrifice restoration quality significantly.}
   \label{fig:svr_general}\vspace{-1em}
\end{figure*}

The general framework of SVR-based restoration can be summarized in Fig.~\ref{fig:svr_general}. An input image $X\!\in\!\mathbb{R}^{w\times h\times c}$ is  embedded into a latent space as latent feature $Y\!\in\!\mathbb{R}^{u \times v\times d}$ by an embedding network $E^{emb}$. Using a learned codebook $\mathcal{C}\!=\!\{c_l\!\in\!\mathbb{R}^d\}$ as in Fig.~\ref{fig:svr_general}~(a), the latent $Y$ is further mapped into a discrete quantized latent feature $Y^q\!\in\!\mathbb{R}^{u \times v\times d}$. Each super-pixel $y^q(l)$ ($l\!=\!1,\ldots,\!u\times\!v$) in $Y^q$ corresponds to a codeword $c_l\!\in\!\mathcal{C}$ that is closest to the corresponding latent feature $y(l)$ in $Y$: \vspace{-2em}

\begin{equation}
    y^q(l)=c_l= \mathit{argmin}_{c_i\in\mathcal{C}} D(c_i,y(l))). \nonumber
\end{equation}
$y^q(l)$ can be represented by the index $z_l$ of codeword $c_l$, and the entire $Y^q$ can be mapped to an $n$-dim vector $Z$ of integers, $n\!=\!u\!\times\!v$. Based on indices $Z$, the quantized feature $Y^q$ can be retrieved from the codebook $\mathcal{C}$. Also, multi-scale features $F$ are computed from several downsampling blocks in $E^{emb}$, which are fed to the corresponding upsampling blocks in a reconstruction network $E^{rec}$ as residual inputs. $E^{rec}$ then reconstructs the output image $\hat{x}$ based on the quantized latent $Y^q$ and features $F$. 

To improve the performance for general image restoration, in Fig.~\ref{fig:svr_general} (b) instead of using one codebook as in \cite{MAGE,CodeFormer2022}, AdaCode \cite{AdaCode} learns a set of basis codebooks $\mathcal{C}_1,
\ldots,\mathcal{C}_{K}$, each corresponding to a semantic partition of the latent space. A weight map $W\!\in\!\mathbb{R}^{u\times v\times K}$ is computed to combine the basis codebooks for adaptive restoration. The quantized latent $Y^q$ is a weight combination of individual quantized latents $Y^q_1,\ldots,Y^q_K$ using each of the basis codebooks: 
\begin{equation}
    y^q(l) = \sum\nolimits_{j=1}^{K} w_{j}(l)y^q_{j}(l), \label{Eqn:weightedcombine}
\end{equation}
and $w_j(l)$ is the weight of the $j$-th codebook for the $l$-th super-pixel in $W$.   

For the purpose of compression, previous methods \cite{NTIREFaceCompress,WACVLIC2024} remove the direct skip connections of the multi-scale features $F$ that are too heavy to transfer. Only the indices $Z_1,\ldots,Z_K$ are sent to the decoder to retrieve the quantized latent $Y^q$ for reconstruction. However, the multi-scale features $F$ provide important fidelity details of the input, and without $F$ the reconstructed result may lack details and may not be consistent with the original input.

\begin{figure*}[t] 
  \centering
   \includegraphics[width=\linewidth]{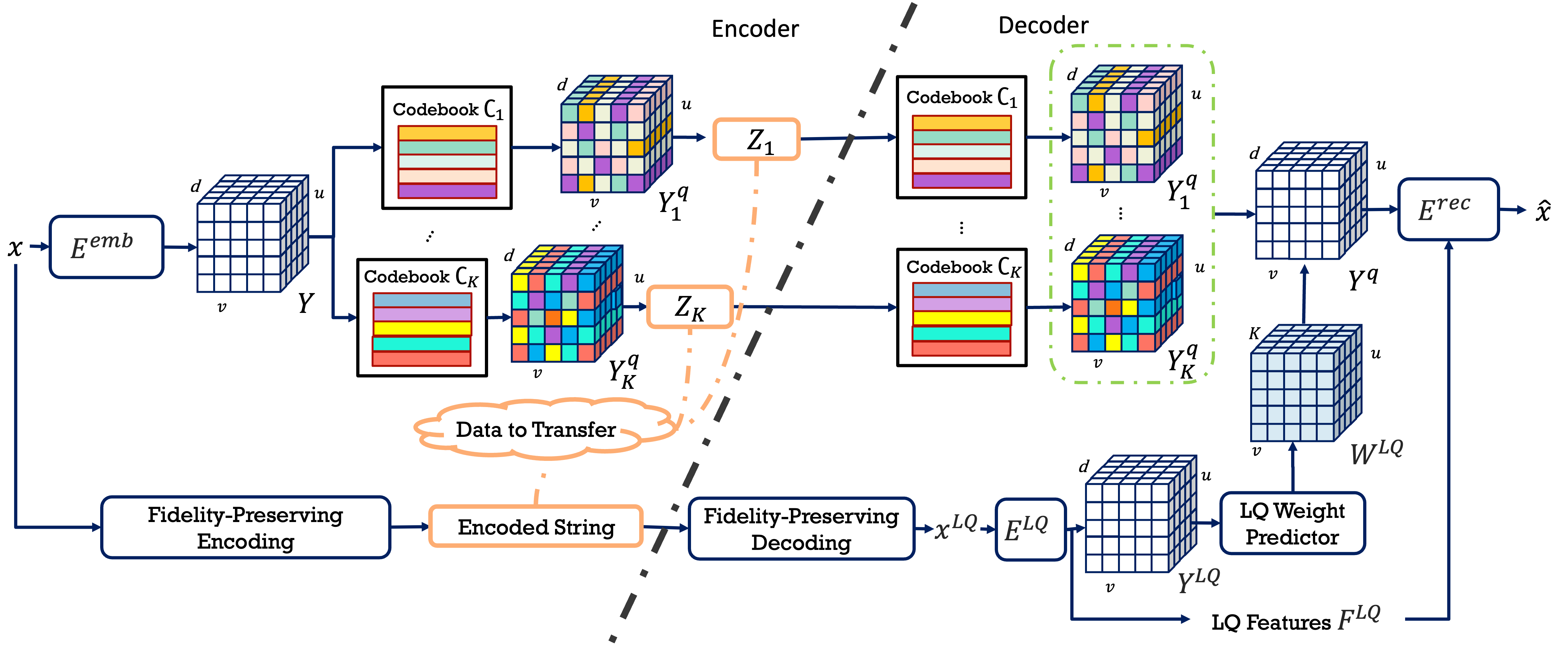}\vspace{-1em}
   \centering{(a) Image compression pipeline}\vspace{.5em}
   \includegraphics[width=\linewidth]{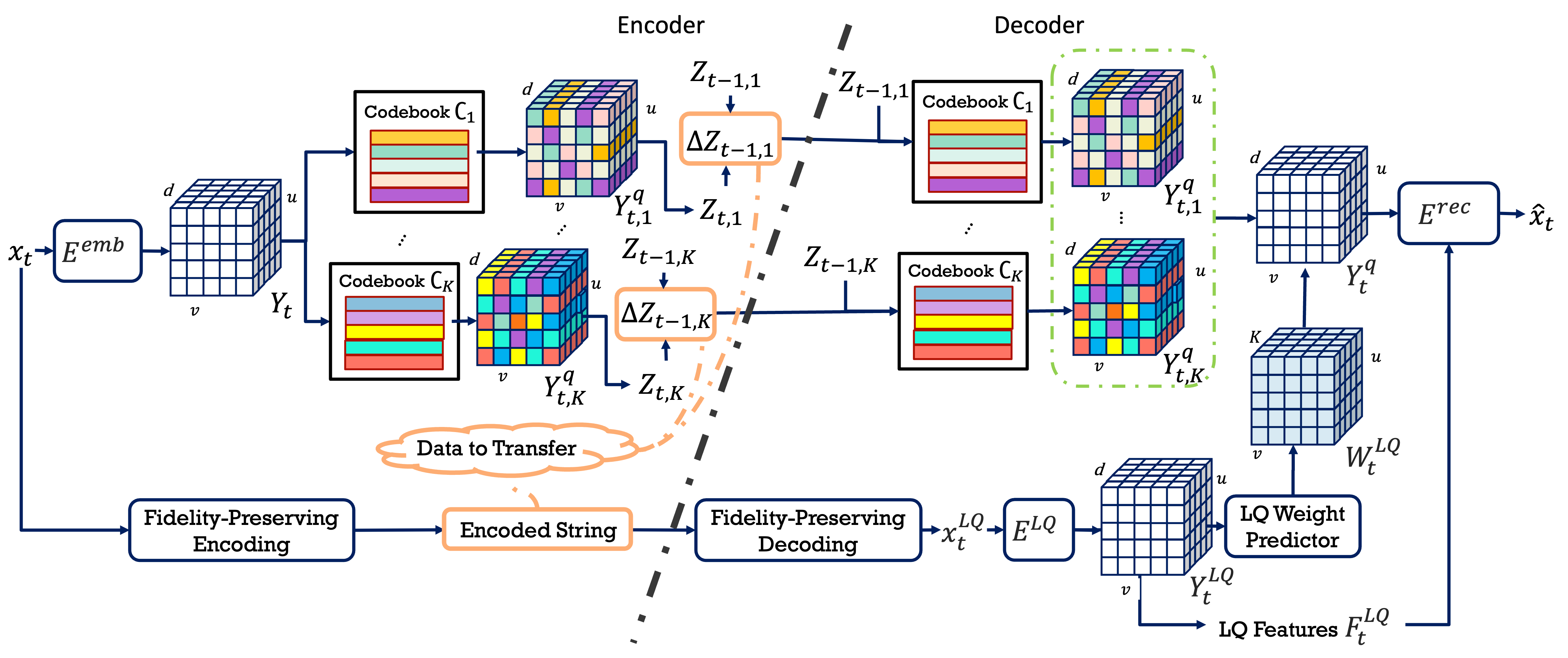}\vspace{-1em}
   \centering{(b)Video compression pipeline} \vspace{-1em}
   \caption{The proposed SVR-based compression framework using LQ control conditions.}
   \label{fig:svr-lq-control}\vspace{-1.5em}
\end{figure*}

\subsection{SVR-based Compression using LQ Control Conditions}

In the realm of image generation, ControlNet \cite{ControlNet2023} has been developed to enable different levels of control over generated results. The key idea is to provide data-specific conditions to a pre-trained generative model to control the generation process. This is analogous to SVR-based compression, where the reconstruction network $E^{rec}$ generates the output based on quantized feature $Y^q$, and the multi-scale features $F$ provide additional control conditions drawn from the current input. This perspective motivates our compression framework in Fig.~\ref{fig:svr-lq-control}. When used as control conditions, the multi-scale features do not have to come from the original input, and therefore we can avoid transmitting the heavy $F$. Instead, they can be drawn from an LQ substitute of the input $x^{LQ}$ in the decoder, and the LQ substitute can be computed in decoder based on fidelity-preserving information calculated by existing compression methods like \cite{vvc_std,MLIC2023} with high compression rates and low bitrate. With the help of additional controls, this framework not only improves the restoration fidelity and quality, but also enables flexible quality control. By tuning the bitrate of the LQ substitute, we can tune the quality of the LQ substitute and change the quality of the control condition. 

To simplify our description in the remaining of this section, without loss of generality, we adopt the multi-codebook notation from Fig.~\ref{fig:svr_general} (b) to describe our approach. For the case where a single codebook is used, we only have $K\!=\!1$ codebook, and there is no need to compute a predicted weight map. \vspace{-.5em}

\subsubsection{SVR-based LIC}

As shown in Fig.~\ref{fig:svr-lq-control} (a), in the encoder, the input image $x$ is embedded into the latent space as latent feature $Y$, which is further quantized into  $Y^q_1,\ldots,Y^q_K$ with associated codeword indices $Z_1,\ldots,Z_K$, by using codebooks $\mathcal{C}_1,\ldots,\mathcal{C}_{K}$, respectively. At the same time, $x$ is encoded into a highly compressed string with low bitrate using a fidelity-preserving image compression method (\textit{e.g.}, an existing LIC method \cite{MIMLIC}), which is transferred to the decoder together with indices $Z_1,\ldots,Z_K$. Then in the decoder, the quantized latents $Y^q_1,\ldots,Y^q_K$ are retrieved from the corresponding codebooks using the codeword indices, and the LQ substitute $x^{LQ}$ is decoded from the corresponding image compression method. Albeit low quality, $x^{LQ}$ carries important fidelity information about the original $x$ to guide reconstruction. 

In the decoder, $x^{LQ}$ is fed into an LQ embedding network $E^{LQ}$ to compute the multi-scale LQ features $F^{LQ}$ from the multiple downsampling blocks in $E^{LQ}$, where $F^{LQ}$ has the same size as the original multi-scale features $F$ (if computed from the original $E^{emb}$). When $K\!>\!1$, an LQ weight map $W^{LQ}\!\in\!\mathbb{R}^{u\times v\times K}$ is also computed by an LQ weight predictor. Both $F^{LQ}$ and $W^{LQ}$ are control conditions drawn from the fidelity-preserving LQ substitute $x^{LQ}$, which are used by the reconstruction network $E^{rec}$ to guide the reconstruction process. Since $F^{LQ}$ and $W^{LQ}$ are calculated in decoder using $x^{LQ}$, they do not increase bitrate.

In detail, to reconstruct the output, when $K\!>\!1$, the final quantized latent $Y^{q}$ is a weighted combination of $Y^q_1,\ldots,Y^q_K$ similar to Eqn.~(\ref{Eqn:weightedcombine}), using LQ weight $W^{LQ}$. When $K\!=\!1$, $Y^q_1\!=\!Y^q$. Then the multi-scale LQ feature $F^{LQ}$ is used as modulating conditions to the multiple upsampling blocks in $E^{rec}$ to guide the reconstruction from $Y^q$. In this work, we use the Controllable Feature Transformation (CFT) module from \cite{CodeFormer2022} to apply modulating conditions. Let $\Theta_{c}$ denote the parameters of the CFT module $CFT_{code}$ to combine the codebook-based quantized feature $Y^{q}$ and the LQ control feature $F^{LQ}$. $F^{LQ}$ tunes $Y^{q}$ into a modulated $Y^{mod}\!=\!Y^{q}\!+\!\alpha_c\!*\!(\beta_c\!*\!Y^{q}\!+\!\gamma_c)$, where $\beta_c,\gamma_c$ are affine parameters $\beta_c,\gamma_c\!=\!\Theta_c(concat(Y^{q},F^{LQ}))$, and $concat(\cdot)$ is the concatenation operation. $\alpha_c$ determines the strength of the control feature $F^{LQ}$ in conditioning the codebook-based feature $Y^{q}$, which can be flexibly set according to the actual compression needs, \textit{i.e.}, to pursue high perceptual quality or high fidelity.

One advantage of our SVR-based LIC method is the flexibility in accommodating different scenarios. For homogeneous computing platforms, our method combines the strength of the fidelity cue from existing fidelity-preserving image compression methods (classic or learning-based) and the perceptual cue from SVR-based restoration, enables bitrate control by tuning the bitrate of the LQ substitute, and allows tradeoff between perceptual quality and fidelity. For heterogeneous computing platforms where previous LIC methods may have difficulty to apply, our method can still give a decent low-bitrate baseline reconstruction with good perceptual quality using SVR-based restoration alone, or can pair with classic compression methods for improved reconstruction.\vspace{-.5em}

\subsubsection{SVR-based LVC}

The above SVR-based LIC method can be easily extended to an effective SVR-based LVC method, whose workflow is shown in Fig.~\ref{fig:svr-lq-control}~(b). Since SVR is robust to input degradation and perturbations, for most frames in a video, a large portion of the codeword indices can be the same between adjacent frames. Therefore, for a video frame $x_t$ at time stamp $t\!>\!1$, only the different indices $\Delta Z_{t,1},\ldots,\Delta Z_{t,K}$ from the previous frame need to be transmitted for the decoder to restore the quantized latent $Y^q_t$ for SVR-based reconstruction. Actually our experiments show that 47\% of the codeword indices remain unchanged on average, leading to effective bit reduction for LVC. The remaining processing modules are similar to the LIC method, with the difference that the LQ substitute $x^{LQ}_t$ of frame $x_t$ comes from a fidelity-preserving video compression method (\textit{e.g.}, classic VVC \cite{vvc_std} or learning based DVC \cite{DVC2019}) or a fidelity-preserving image compression method. 

Similar to SVR-based LIC, our SVR-based LVC method provides flexibility to accommodate different scenarios, where we can choose different methods to generate the LQ substitutes by considering different factors like computation and transmission requirements, reconstruction targets, \textit{etc.} In addition, there is no error propagation in recovering the codeword-based quantized feature for every frame, and a decent low-bitrate baseline with good perceptual quality can be mostly guaranteed. Furthermore,  in comparison to previous LIC and LVC methods that usually have completely different processing pipelines, the SVR-based LIC and LVC have a similar workflow with similar processing modules, making it possible to simplify industrial productive optimization. 

It is worth mentioning that in our implementation the reconstruction network $E^{rec}$ has the same architecture for both LIC and LVC. A video-oriented network like C3D \cite{C3D2015} can be used for LVC to better ensure temporal consistency. We found it unnecessary in experiments since our result is quite consistent temporally due to the deterministic generation and temporal-consistent fidelity control.\vspace{-.5em} 

\subsubsection{Complexity}
Our SVR-based LIC and LVC are quite efficient in computation. The main SVR-based restoration only involves inference through the embedding network $E^{emb}$ in encoder, and through the LQ embedding network $E^{LQ}$ and reconstruction network $E^{rec}$ in decoder. For fidelity-preserving control, only LQ substitute $x^{LQ}$ is needed, and we can get fast computation by using high compression rates, \textit{e.g.}, by downsampling the input and turning off unnecessary processing modules in fidelity-preserving image/video compression tools. 

In comparison, besides going through the VAE encoding and decoding networks, previous hyperprior-based LIC methods usually need an auto-regressive computation to encode and decode accurate hyperpriors in CPU, which can be very time consuming. Previous LVC methods have very high computation costs in general, requiring not only inference through multiple networks in different modules but also the entire decoding computation in  encoder to obtain residues.

\subsection{Training strategy}

We train our models in three different stages. \vspace{-.5em}

\subsubsection{Pretrain} The embedding network $E^{emb}$, the codebooks $\mathcal{C}_1,\ldots,\mathcal{C}_{K}$, the reconstruction network $E^{rec}$, and the weight predictor (when $K\!>\!1$) are pretrained for single-codebook-based image restoration \cite{FeMaSR} or multi-codebook-based image restoration \cite{AdaCode}.\vspace{-.5em}

\subsubsection{Train SVR-based LIC} The embedding network $E^{emb}$ and codebooks $\mathcal{C}_1,\ldots,\mathcal{C}_{K}$ are fixed, and we train the LQ embedding network $E^{LQ}$, the CFT module $CFT_{code}$, the LQ weight predictor (for $K\!>\!1$), the reconstruction network $E^{rec}$, and the GAN discriminator for the LIC pipeline. The training loss comprises of pixel-level $L_1$ loss and SSIM, the perceptual loss \cite{perceptualloss}, LPIPS \cite{lpips}, and the GAN adversarial loss \cite{ganloss}. The straight-through gradient estimation is used for back-propagation through the non-differentiable vector quantization process during training. The strength of control is set as $\alpha_{c}\!=\!1$ for all inputs.\vspace{-.5em}

\subsubsection{Train SVR-based LVC} The embedding network $E^{emb}$, the codebooks $\mathcal{C}_1,\ldots,\mathcal{C}_{K}$, and the reconstruction network $E^{rec}$ are fixed, and we train the LQ embedding network $E^{LQ}$, the CFT module $CFT_{code}$, the LQ weight predictor (when $K\!>\!1$), and the GAN discriminator by finetuning from the corresponding LIC version in the previous stage. One benefit of funetuning from the LIC counterparts is to benefit from the large variety of image training content to avoid overfitting, due to the limited amount of training videos from the standardization community.\vspace{-.5em}

\subsection{Bit reduction for integer codeword indices} \label{sec:bitreduction}

To transfer codeword indices, naively we need $b(Z_k)\!=u\!\times\!v\!\times\!\text{floor}(\log_2 n_k)$ bits for each codebook $\mathcal{C}_k$ of size $n_k$. This number can be further reduced to save bit consumption of the whole system. We propose an effective arithmetic coding method that can losslessly compress the integer indices by $5\!\times$ on average. For natural images, codewords normally show up with different frequencies. For instance, codewords of natural scenes may be used more frequently than those of human faces. We can assign less bits to more frequently used indices to reduce the total bitrate. Specifically, we first calculate the frequency of codewords' usage in training data and reorder the codewords in descending order. Then for each particular indices string of each datum, we convert each odd index $ix$ to a negative integer as $ix^{*}\!=\!-(ix\!+\!1)/2$ and rescale even indices by $1/2$. Such operations transform the indices distribution to a Gaussian style bell shape, which can be efficiently encoded by Gaussian Mixture-based arithmetic coding \cite{arithmeticcoding}. 

\section{Experiments}

\subsubsection{Datasets} We tested the proposed SVR-based LIC and LVC method, respectively, over the JPEG-AI dataset \cite{JPEG-AI-data,JPEGAI} and a mixed video dataset combining test video sequences from several video compression standards including AOM \cite{aom_std}, MPEG \cite{mpeg_std}, JVET \cite{Jvet_std}, and AVS \cite{avs_std}. The JPEG-AI dataset had 5664 images with a large variety of visual content and resolutions up to 8K. The training, validation, and test set had 5264, 350, and 50 images, respectively. This dataset was developed by the JPEG standardization organization to provide a standard benchmark for evaluating LIC methods in the field. The mixed video set contained 150 video sequences, which were used as test sequences by the standardization community to test the video compression algorithms in the field. Here we removed the duplicate sequences, \textit{e.g.}, the same sequences used by different standards, or the same sequences resized to different resolutions, where we kept videos with different resolutions ranging from $240\!\times\!400$ to 4K. 134 and 16 video sequences were used for training and test respectively.

The training patches had $256\!\times\!256$ resolution, which were randomly cropped from randomly resized training images or video frames, and were augmented by random flipping and rotation. For evaluation, the maximum resolution of inference tiles was $1080\!\times\!1080$. For training SVR-based LVC modules, video frames were randomly sampled from videos, and were used as images in the same way as training the SVR-based LIC modules. For all tested methods, each training stage had 500K iterations with Adam optimizer and a batch size of 32, using 8 NVIDIA Tesla V100 GPUs. The learning rate for the generator and discriminator were fixed as 1e-4 and 4e-4, respectively. 

\subsubsection{Evaluation Metrics} For reconstruction distortion, we measured PSNR and SSIM, as well as the perceptual LPIPS \cite{lpips}. The bitrate was measured by bpp (bit-per-pixel): $bpp\!=\!B/(h\!\times\!w)$, and the overall bits $B\!=\!b_c\!+\!b_{LQ}$ consisted of $b_c$ for sending codebook indices and $b_{LQ}$ for sending the encoded string to compute the LQ substitute $x^{LQ}$ using previous image/video compression methods. In detail, for LIC $b_c=\sum_{k=1}^K b(Z_k)$, and for LVC $b_c=\sum_{k=1}^K[b(Z_{1,k})+\sum_{t=2}^Tb(\Delta Z_{t,k})]/T$.  $b(Z_k)$ is computed based on the indices reduction method described in Sec.~\ref{sec:bitreduction}. In terms of $b_{LQ}$, it determined the quality of the LQ substitute $x^{LQ}$. We chose a low $b_{LQ}$ ($<0.1$ bpp) to roughly match $b_c$. \vspace{-.5em}

\begin{figure*}[t]
    \centering
    \begin{minipage}[b][0.325\linewidth][b]{0.325\linewidth}
    \includegraphics[width=\textwidth]{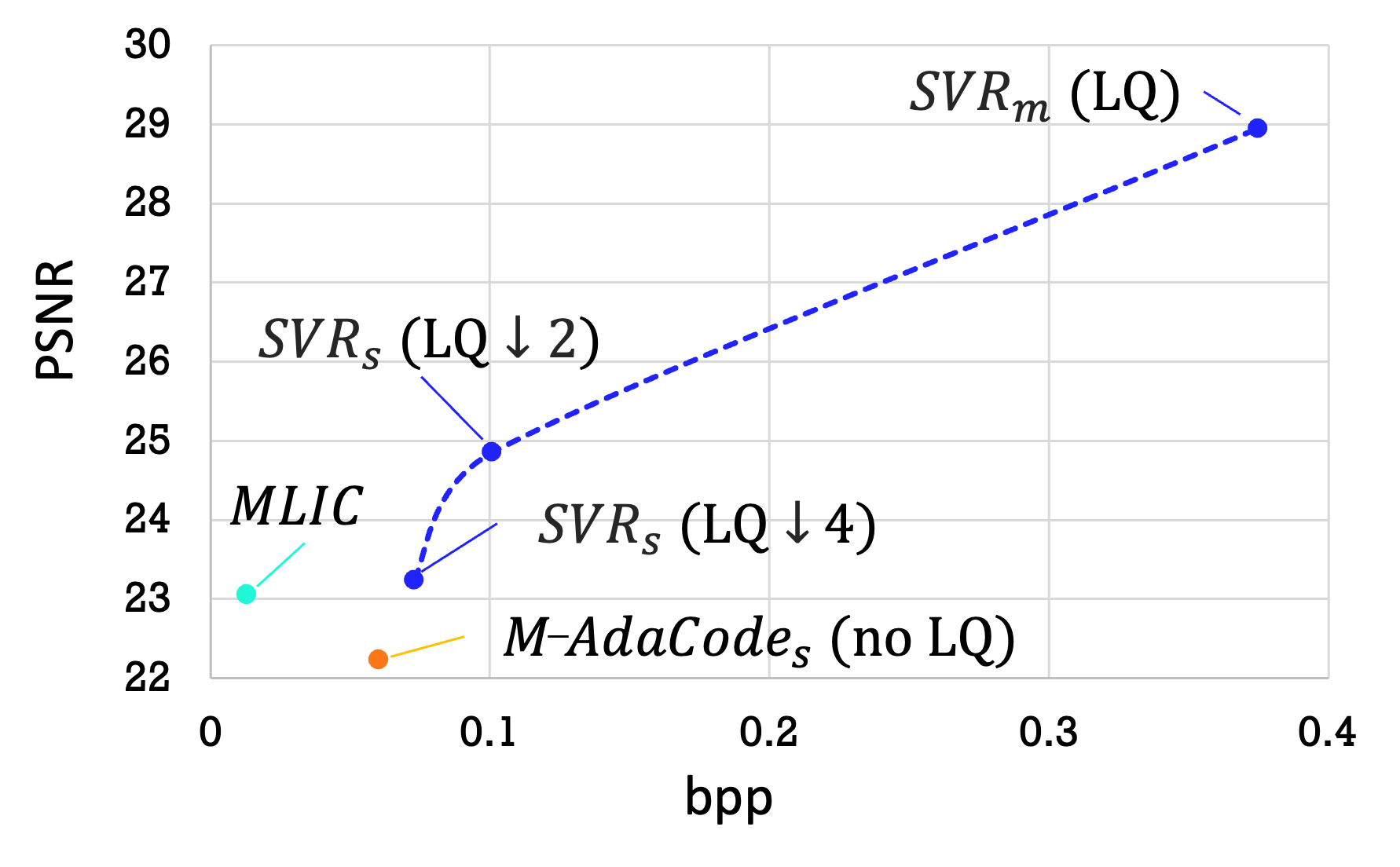}
    \end{minipage}
    \begin{minipage}[b][0.325\linewidth][b]{0.325\linewidth}
    \includegraphics[width=\textwidth]{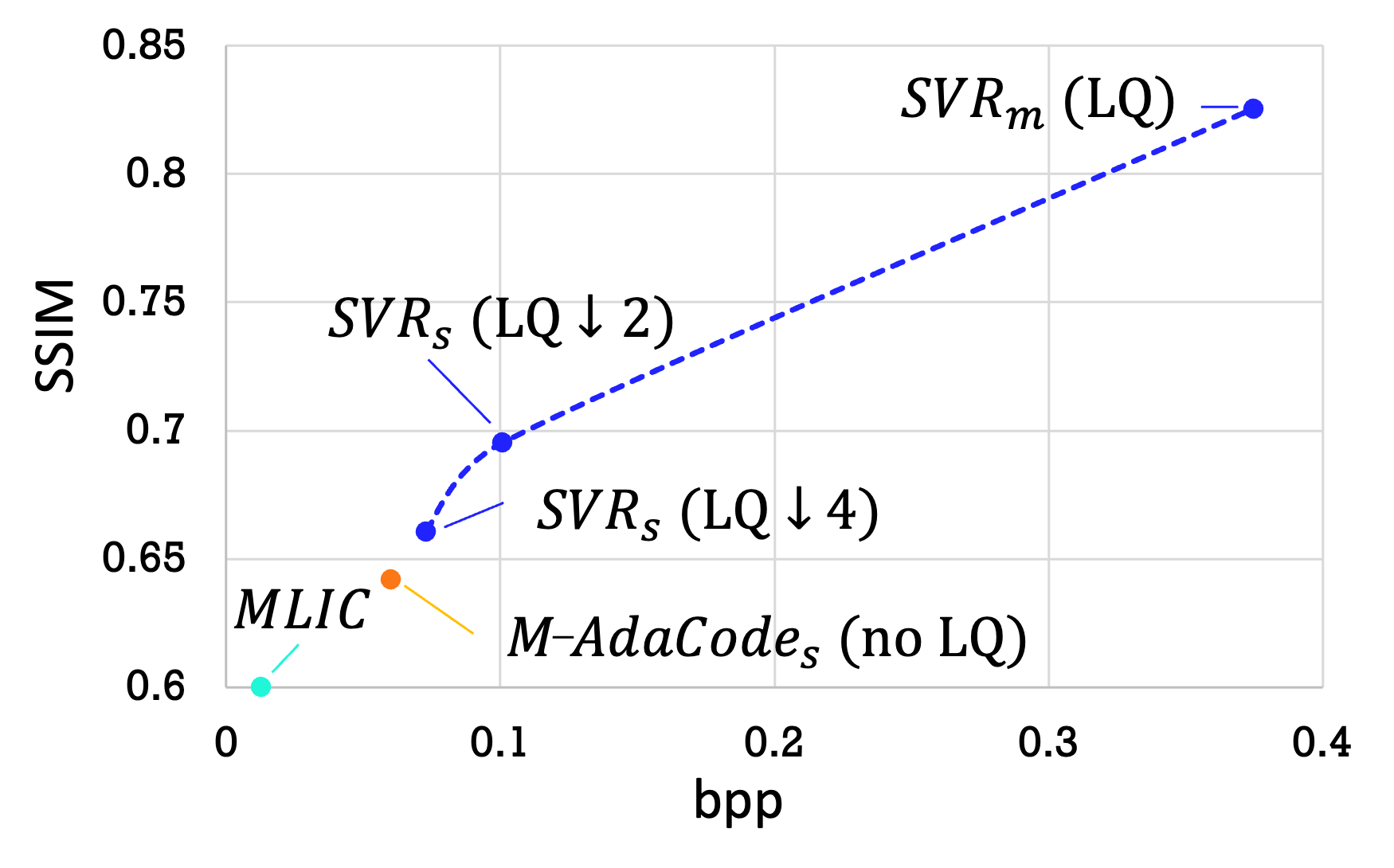}
    \end{minipage}
    \begin{minipage}[b][0.325\linewidth][b]{0.325\linewidth}
    \includegraphics[width=\textwidth]{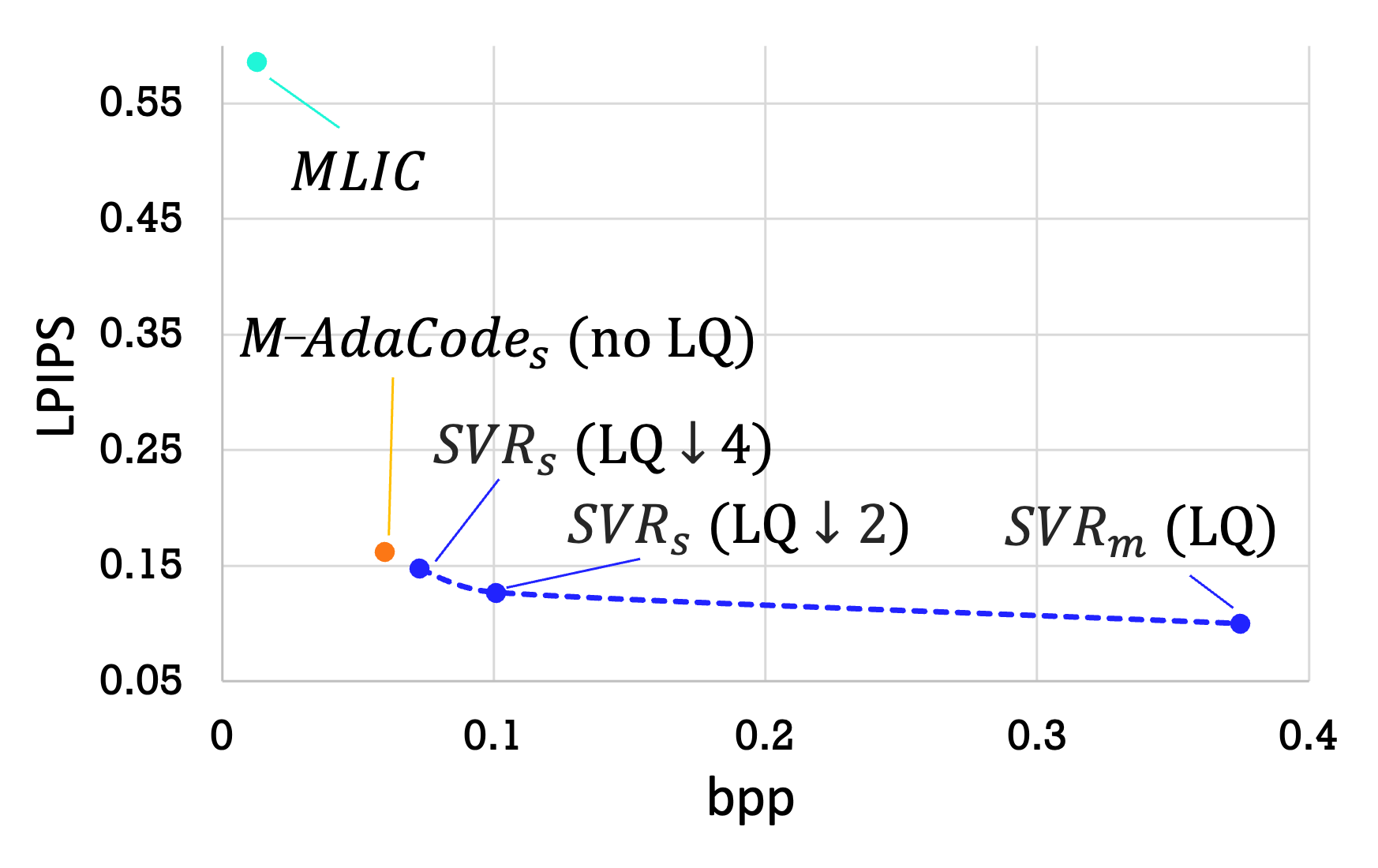}
    \end{minipage}    \vspace{-1em}
\caption{Rate-distortion performance for image compression.}
\label{fig:LIC-performance}\vspace{-1em}
\end{figure*}

\subsubsection{Evaluated Methods}

We evaluated two configurations for our approach using SOTA SVR-based restoration algorithms: the single-codebook-based FeMaSR \cite{FeMaSR} and the multi-codebook-based AdaCode \cite{AdaCode}. With only a single codebook, FeMaSR gave very low bitrate. Using multiple codebooks, AdaCode gave improved restoration quality but consumed more bits. 

To generate fidelity-preserving  LQ substitute $x^{LQ}$, for SVR-based LIC we used previous SOTA LIC method MLIC \cite{MLIC2023}. The pre-trained MLIC model with the lowest available bitrate setting was used, which corresponded to the quality-1 model in \cite{MLIC2023}. In order to get lower bitrate for $b_{LQ}$ to match $b_c$, we first downsampled the input $x$ by $2\!\times$ or $4\!\times$, and then used MLIC to encode the downsampled input and then upsampled the decoded $x^{LQ}$ back to the original size. The bicubic filter was used for  downsampling/upsampling. For SVR-based LVC we used the SOTA VVC video compression method \cite{vvc_std} with $qp\!=\!42$ to generate $x^{LQ}$, which gave reasonable low-bitrate reconstruction in general.\vspace{-.5em}

\subsection{LIC Results}

Fig.~\ref{fig:LIC-performance} gives the rate-distortion performance for image compression. For our SVR-based LIC, we tested 3 different settings: single-codebook SVR with $2\times$ and $4\times$ downsampled-upsampled $x^{LQ}$ as ``$\text{SVR}_s$(LQ$\downarrow\!\!2\times$)'' and ``$\text{SVR}_s$(LQ$\downarrow\!\!4\times$)'', and multi-codebook SVR with $x^{LQ}$ without downsampling-upsampling $\text{SVR}_m(LQ)$. We also compared with M-AdaCode without $x^{LQ}$ with the 1-codebook setting \cite{WACVLIC2024} (``$\text{M-AdaCode}_s$(no LQ)'') and compared with MLIC \cite{MLIC2023} generated $x^{LQ}$. From the figures,  ``$\text{SVR}_s$(LQ$\downarrow\!\!4\times$)'' outperformed M-AdaCode with 1\textit{dB}, 2.3\% and 8.9\% improvements over PSNR, SSIM and LPIPS, respectively, using only a 0.013\textit{bpp} increase.  Compared to MLIC, ``$\text{SVR}_s$(LQ$\downarrow\!\!4\times$)'' improved LPIPS and SSIM by 74.8\%, and 10.1\% respectively. Among methods having $<\!0.1 bpp$, our SVR-based LIC gave balanced results with good fidelity and perceptual quality.

Fig.~\ref{fig:LIC-example} gives some restoration examples, which clearly show the strength of our method. With the help of $x^{LQ}$, our SVR-based LIC  largely improved the reconstruction fidelity and perceptual quality  with rich visually pleasing details, compared to $\text{M-AdaCode}_s$(no LQ)''. Also, our framework can be flexibly configured to different settings to tradeoff bitrate and reconstruction quality. \vspace{-.5em}

\subsection{LVC Results}

\begin{figure*}[t]

    \centering
    \begin{minipage}[b][0.13\linewidth][b]{0.2\linewidth}
    \centering
    \centerline{\scriptsize{\textcolor{white}{g}$4096\!\times\!2160$\textcolor{white}{g}}}
    \vspace{0.03cm}
    \includegraphics[width=\textwidth,height=0.645\linewidth]{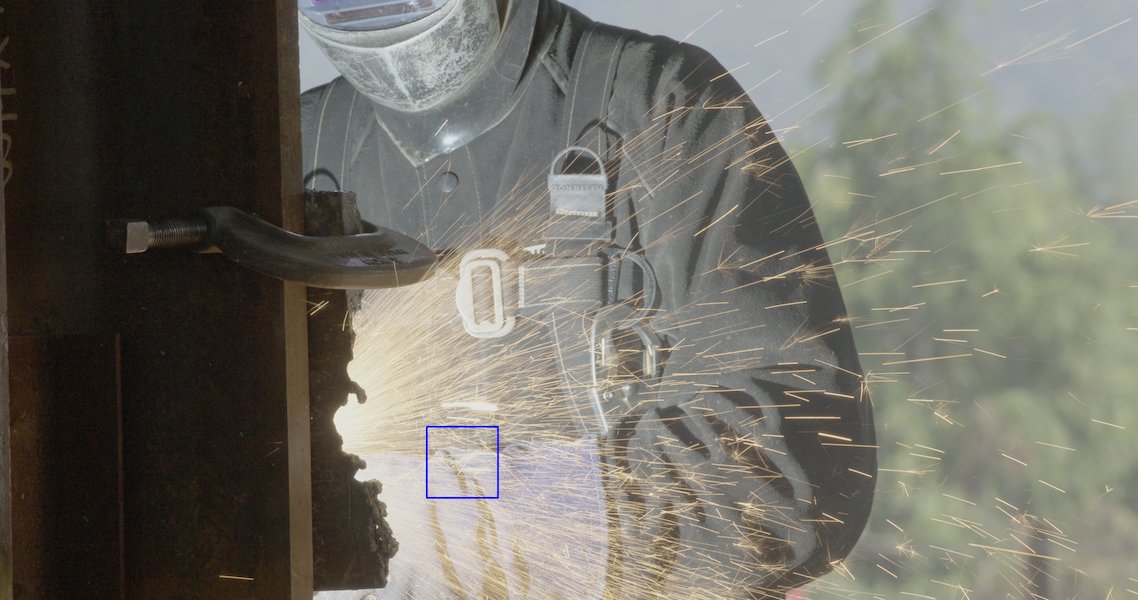}
    \vspace{-0.3cm}
    \centerline{\tiny\textcolor{white}{g}{$86\%$ index change}}\medskip
    \end{minipage}
    \hspace{0.15cm}
    \begin{minipage}[b]{0.13\linewidth}
    \centering
    \centerline{\scriptsize \textcolor{white}{g} \textcolor{white}{g}}
    \vspace{0.02cm}
    \includegraphics[width=\textwidth]{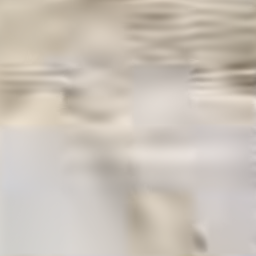}
    \vspace{-0.3cm}
    \centerline{\tiny 0.534$\mid$30.51$\mid$0.825}\medskip
    \end{minipage}
    \hspace{0.15cm}
    \begin{minipage}[b]{0.13\linewidth}
    \centering
    \centerline{\scriptsize \textcolor{white}{g} \textcolor{white}{g}}
    \vspace{0.02cm}
    \includegraphics[width=\textwidth]{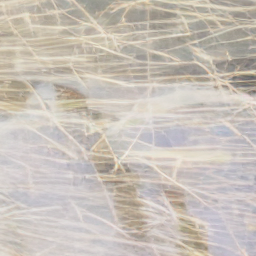}
    \vspace{-0.3cm}
    \centerline{\tiny 0.149$\mid$32.48$\mid$0.858}\medskip
    \end{minipage}
    \hspace{0.15cm}
    \begin{minipage}[b]{0.13\linewidth}
    \centering
    \centerline{\scriptsize  \textcolor{white}{g} \textcolor{white}{g}}
    \vspace{0.02cm}
    \includegraphics[width=\textwidth]{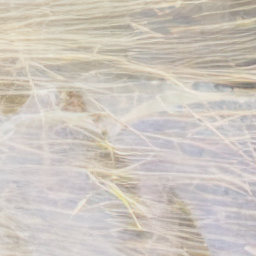}
    \vspace{-0.3cm}
    \centerline{\tiny 0.152$\mid${32.02}$\mid$0.857}\medskip
    \end{minipage}
    \hspace{0.15cm}
    \begin{minipage}[b]{0.13\linewidth}
    \centering
    \centerline{\scriptsize \textcolor{white}{g} \textcolor{white}{g}}
    \vspace{0.02cm}
    \includegraphics[width=\textwidth]{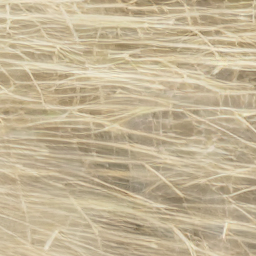}
    \vspace{-0.3cm}
    \centerline{\tiny {0.160}$\mid$29.79$\mid$0.834}\medskip
    \end{minipage}
    \hspace{0.15cm}
    \begin{minipage}[b]{0.13\linewidth}
    \centering
    \centerline{\scriptsize \textcolor{white}{g} \textcolor{white}{g}}
    \vspace{0.02cm}
    \includegraphics[width=\textwidth]{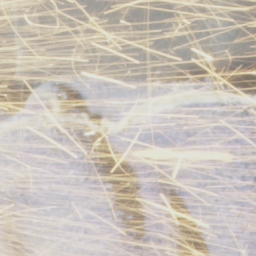}
    \vspace{-0.3cm}
    \centerline{\tiny\textcolor{white}{0.413$\mid$}}\medskip
    \end{minipage}    
    \vspace{.6em}

    \centering
    \begin{minipage}[b][0.13\linewidth][b]{0.2\linewidth}
    \centering
    \centerline{\scriptsize{\textcolor{white}{g}$2560\!\times\!1440$\textcolor{white}{g}}}
    \vspace{0.03cm}
    \includegraphics[width=\textwidth,height=0.645\linewidth]{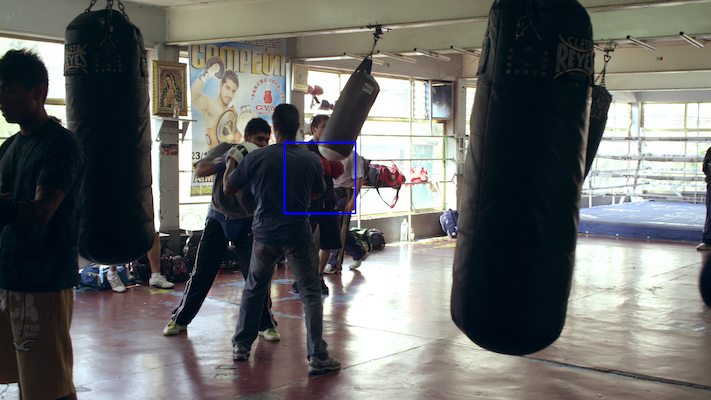}
    \vspace{-0.3cm}
    \centerline{\tiny\textcolor{white}{g}{$72\%$ index change}}\medskip
    \end{minipage}
    \hspace{0.15cm}
    \begin{minipage}[b]{0.13\linewidth}
    \centering
    \centerline{\scriptsize \textcolor{white}{g} \textcolor{white}{g}}
    \vspace{0.02cm}
    \includegraphics[width=\textwidth]{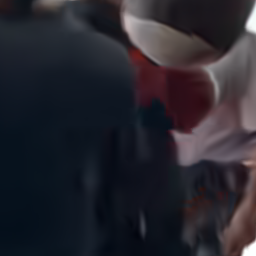}
    \vspace{-0.3cm}
    \centerline{\tiny 0.271$\mid$32.78$\mid$0.904}\medskip
    \end{minipage}
    \hspace{0.15cm}
    \begin{minipage}[b]{0.13\linewidth}
    \centering
    \centerline{\scriptsize \textcolor{white}{g} \textcolor{white}{g}}
    \vspace{0.02cm}
    \includegraphics[width=\textwidth]{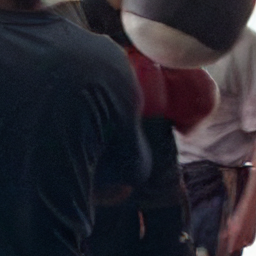}
    \vspace{-0.3cm}
    \centerline{\tiny 0.084$\mid$32.62$\mid$0.886}\medskip
    \end{minipage}
    \hspace{0.15cm}
    \begin{minipage}[b]{0.13\linewidth}
    \centering
    \centerline{\scriptsize  \textcolor{white}{g} \textcolor{white}{g}}
    \vspace{0.02cm}
    \includegraphics[width=\textwidth]{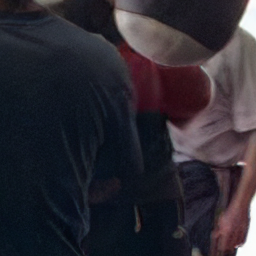}
    \vspace{-0.3cm}
    \centerline{\tiny 0.083$\mid${32.57}$\mid$0.886}\medskip
    \end{minipage}
    \hspace{0.15cm}
    \begin{minipage}[b]{0.13\linewidth}
    \centering
    \centerline{\scriptsize \textcolor{white}{g} \textcolor{white}{g}}
    \vspace{0.02cm}
    \includegraphics[width=\textwidth]{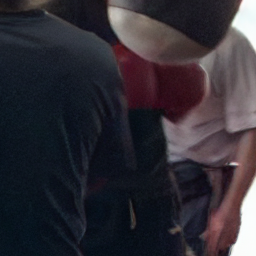}
    \vspace{-0.3cm}
    \centerline{\tiny {0.083}$\mid$32.81$\mid$0.888}\medskip
    \end{minipage}
    \hspace{0.15cm}
    \begin{minipage}[b]{0.13\linewidth}
    \centering
    \centerline{\scriptsize \textcolor{white}{g} \textcolor{white}{g}}
    \vspace{0.02cm}
    \includegraphics[width=\textwidth]{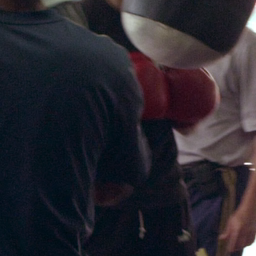}
    \vspace{-0.3cm}
    \centerline{\tiny\textcolor{white}{0.413$\mid$}}\medskip
    \end{minipage}    
    \vspace{.6em}   
    
    \centering
    \begin{minipage}[b][0.13\linewidth][b]{0.2\linewidth}
    \centering
    \centerline{\scriptsize{\textcolor{white}{g}$1920\!\times\!1080$\textcolor{white}{g}}}
    \vspace{0.03cm}
    \includegraphics[width=\textwidth,height=0.645\linewidth]{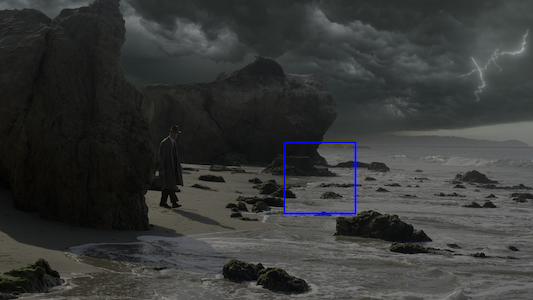}
    \vspace{-0.3cm}
    \centerline{\tiny\textcolor{white}{g}{$46\%$ index change}}\medskip
    \end{minipage}
    \hspace{0.15cm}
    \begin{minipage}[b]{0.13\linewidth}
    \centering
    \centerline{\scriptsize \textcolor{white}{g} \textcolor{white}{g}}
    \vspace{0.02cm}
    \includegraphics[width=\textwidth]{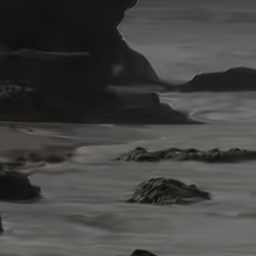}
    \vspace{-0.3cm}
    \centerline{\tiny 0.423$\mid$32.30$\mid$0.843}\medskip
    \end{minipage}
    \hspace{0.15cm}
    \begin{minipage}[b]{0.13\linewidth}
    \centering
    \centerline{\scriptsize \textcolor{white}{g} \textcolor{white}{g}}
    \vspace{0.02cm}
    \includegraphics[width=\textwidth]{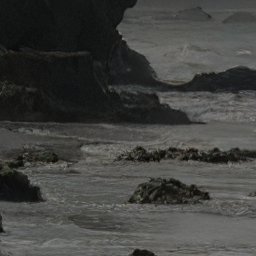}
    \vspace{-0.3cm}
    \centerline{\tiny 0.103$\mid$32.03$\mid$0.826}\medskip
    \end{minipage}
    \hspace{0.15cm}
    \begin{minipage}[b]{0.13\linewidth}
    \centering
    \centerline{\scriptsize  \textcolor{white}{g} \textcolor{white}{g}}
    \vspace{0.02cm}
    \includegraphics[width=\textwidth]{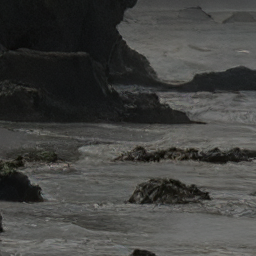}
    \vspace{-0.3cm}
    \centerline{\tiny 0.102$\mid${32.10}$\mid$0.826}\medskip
    \end{minipage}
    \hspace{0.15cm}
    \begin{minipage}[b]{0.13\linewidth}
    \centering
    \centerline{\scriptsize \textcolor{white}{g} \textcolor{white}{g}}
    \vspace{0.02cm}
    \includegraphics[width=\textwidth]{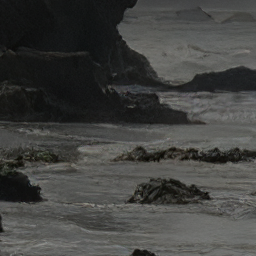}
    \vspace{-0.3cm}
    \centerline{\tiny {0.099}$\mid$32.23$\mid$0.828}\medskip
    \end{minipage}
    \hspace{0.15cm}
    \begin{minipage}[b]{0.13\linewidth}
    \centering
    \centerline{\scriptsize \textcolor{white}{g} \textcolor{white}{g}}
    \vspace{0.02cm}
    \includegraphics[width=\textwidth]{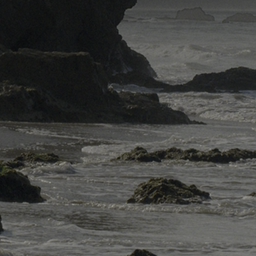}
    \vspace{-0.3cm}
    \centerline{\tiny\textcolor{white}{0.413$\mid$}}\medskip
    \end{minipage}    
    \vspace{.6em}    

    \centering
    \begin{minipage}[b][0.13\linewidth][b]{0.2\linewidth}
    \centering
    \centerline{\scriptsize{\textcolor{white}{g}$1920\!\times\!1080$\textcolor{white}{g}}}
    \vspace{0.03cm}
    \includegraphics[width=\textwidth,height=0.645\linewidth]{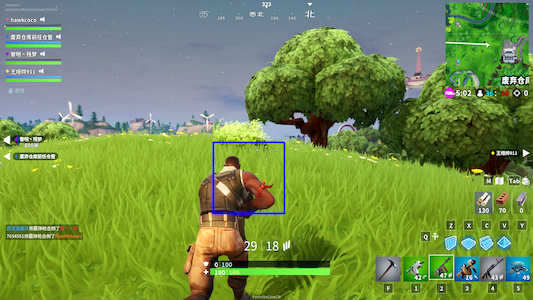}
    \vspace{-0.3cm}
    \centerline{\tiny\textcolor{white}{g}{$87\%$ index change}}\medskip
    \end{minipage}
    \hspace{0.15cm}
    \begin{minipage}[b]{0.13\linewidth}
    \centering
    \centerline{\scriptsize \textcolor{white}{g} \textcolor{white}{g}}
    \vspace{0.02cm}
    \includegraphics[width=\textwidth]{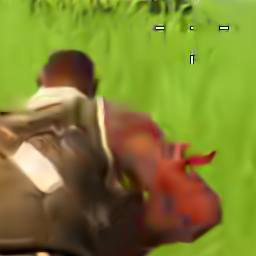}
    \vspace{-0.3cm}
    \centerline{\tiny 0.304$\mid$28.26$\mid$0.812}\medskip
    \end{minipage}
    \hspace{0.15cm}
    \begin{minipage}[b]{0.13\linewidth}
    \centering
    \centerline{\scriptsize \textcolor{white}{g} \textcolor{white}{g}}
    \vspace{0.02cm}
    \includegraphics[width=\textwidth]{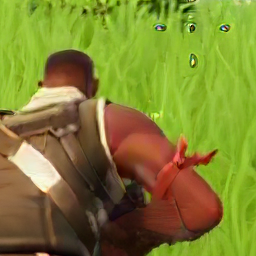}
    \vspace{-0.3cm}
    \centerline{\tiny 0.100$\mid$28.34$\mid$0.834}\medskip
    \end{minipage}
    \hspace{0.15cm}
    \begin{minipage}[b]{0.13\linewidth}
    \centering
    \centerline{\scriptsize  \textcolor{white}{g} \textcolor{white}{g}}
    \vspace{0.02cm}
    \includegraphics[width=\textwidth]{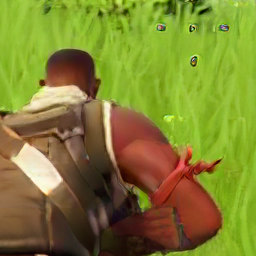}
    \vspace{-0.3cm}
    \centerline{\tiny 0.103$\mid${28.17}$\mid$0.840}\medskip
    \end{minipage}
    \hspace{0.15cm}
    \begin{minipage}[b]{0.13\linewidth}
    \centering
    \centerline{\scriptsize \textcolor{white}{g} \textcolor{white}{g}}
    \vspace{0.02cm}
    \includegraphics[width=\textwidth]{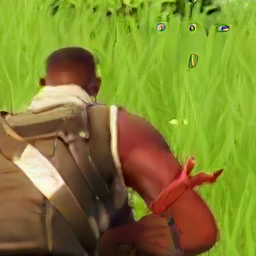}
    \vspace{-0.3cm}
    \centerline{\tiny {0.100}$\mid$27.80$\mid$0.825}\medskip
    \end{minipage}
    \hspace{0.15cm}
    \begin{minipage}[b]{0.13\linewidth}
    \centering
    \centerline{\scriptsize \textcolor{white}{g} \textcolor{white}{g}}
    \vspace{0.02cm}
    \includegraphics[width=\textwidth]{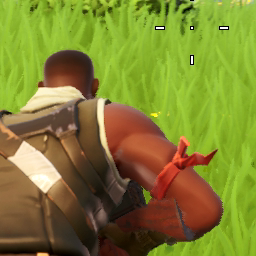}
    \vspace{-0.3cm}
    \centerline{\tiny\textcolor{white}{0.413$\mid$}}\medskip
    \end{minipage}    
    \vspace{.6em}

    \centering
    \begin{minipage}[b][0.13\linewidth][b]{0.2\linewidth}
    \centering
    \centerline{\scriptsize{\textcolor{white}{g}$1280\!\times\!720$\textcolor{white}{g}}}
    \vspace{0.03cm}
    \includegraphics[width=\textwidth,height=0.645\linewidth]{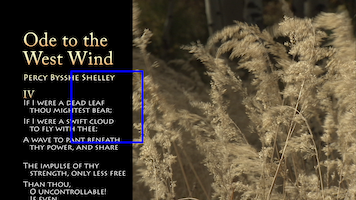}
    \vspace{-0.3cm}
    \centerline{\tiny\textcolor{white}{g}{$61\%$ index change}}\medskip
    \end{minipage}
    \hspace{0.15cm}
    \begin{minipage}[b]{0.13\linewidth}
    \centering
    \centerline{\scriptsize \textcolor{white}{g} \textcolor{white}{g}}
    \vspace{0.02cm}
    \includegraphics[width=\textwidth]{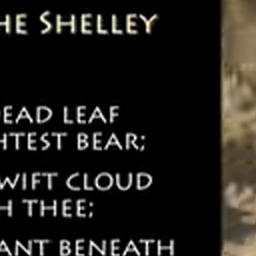}
    \vspace{-0.3cm}
    \centerline{\tiny 0.385$\mid$26.69$\mid$0.759}\medskip
    \end{minipage}
    \hspace{0.15cm}
    \begin{minipage}[b]{0.13\linewidth}
    \centering
    \centerline{\scriptsize \textcolor{white}{g} \textcolor{white}{g}}
    \vspace{0.02cm}
    \includegraphics[width=\textwidth]{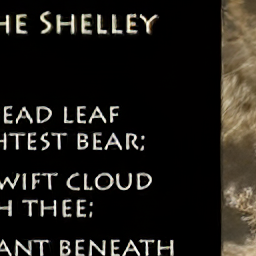}
    \vspace{-0.3cm}
    \centerline{\tiny 0.111$\mid$26.75$\mid$0.782}\medskip
    \end{minipage}
    \hspace{0.15cm}
    \begin{minipage}[b]{0.13\linewidth}
    \centering
    \centerline{\scriptsize  \textcolor{white}{g} \textcolor{white}{g}}
    \vspace{0.02cm}
    \includegraphics[width=\textwidth]{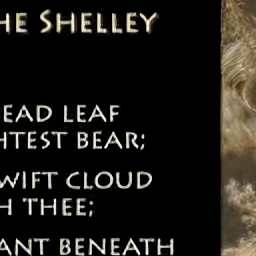}
    \vspace{-0.3cm}
    \centerline{\tiny 0.116$\mid${26.37}$\mid$0.777}\medskip
    \end{minipage}
    \hspace{0.15cm}
    \begin{minipage}[b]{0.13\linewidth}
    \centering
    \centerline{\scriptsize \textcolor{white}{g} \textcolor{white}{g}}
    \vspace{0.02cm}
    \includegraphics[width=\textwidth]{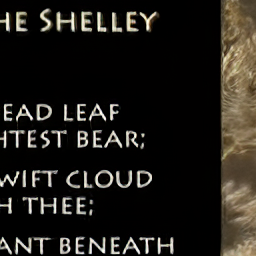}
    \vspace{-0.3cm}
    \centerline{\tiny {0.124}$\mid$26.51$\mid$0.770}\medskip
    \end{minipage}
    \hspace{0.15cm}
    \begin{minipage}[b]{0.13\linewidth}
    \centering
    \centerline{\scriptsize \textcolor{white}{g} \textcolor{white}{g}}
    \vspace{0.02cm}
    \includegraphics[width=\textwidth]{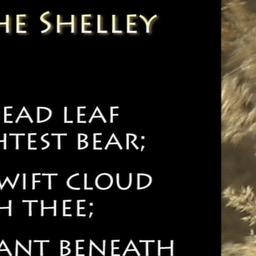}
    \vspace{-0.3cm}
    \centerline{\tiny\textcolor{white}{0.413$\mid$}}\medskip
    \end{minipage}    
    \vspace{.6em}

    \centering
    \begin{minipage}[b][0.13\linewidth][b]{0.2\linewidth}
    \centering
    \centerline{\scriptsize{\textcolor{white}{g}$704\!\times\!480$\textcolor{white}{g}}}
    \vspace{0.03cm}
    \includegraphics[width=\textwidth,height=0.645\linewidth]{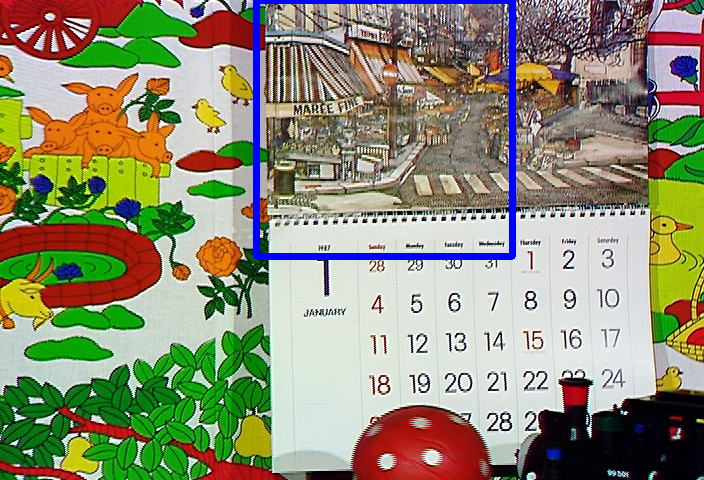}
    \vspace{-0.3cm}
    \centerline{\tiny\textcolor{white}{g}{$43\%$ index change}}\medskip
    \end{minipage}
    \hspace{0.15cm}
    \begin{minipage}[b]{0.13\linewidth}
    \centering
    \centerline{\scriptsize \textcolor{white}{g} \textcolor{white}{g}}
    \vspace{0.02cm}
    \includegraphics[width=\textwidth]{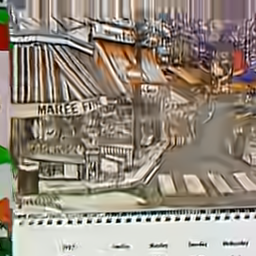}
    \vspace{-0.3cm}
    \centerline{\tiny 0.249$\mid$20.86$\mid$0.713}\medskip
    \end{minipage}
    \hspace{0.15cm}
    \begin{minipage}[b]{0.13\linewidth}
    \centering
    \centerline{\scriptsize \textcolor{white}{g} \textcolor{white}{g}}
    \vspace{0.02cm}
    \includegraphics[width=\textwidth]{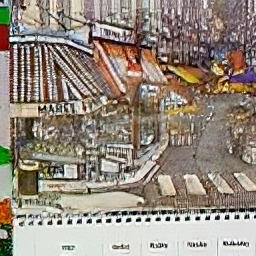}
    \vspace{-0.3cm}
    \centerline{\tiny 0.120$\mid$19.8$\mid$0.678}\medskip
    \end{minipage}
    \hspace{0.15cm}
    \begin{minipage}[b]{0.13\linewidth}
    \centering
    \centerline{\scriptsize  \textcolor{white}{g} \textcolor{white}{g}}
    \vspace{0.02cm}
    \includegraphics[width=\textwidth]{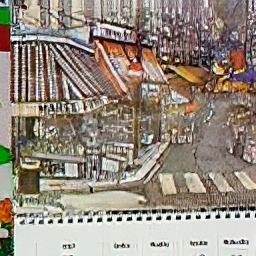}
    \vspace{-0.3cm}
    \centerline{\tiny 0.119$\mid$19.9$\mid$0.692}\medskip
    \end{minipage}
    \hspace{0.15cm}
    \begin{minipage}[b]{0.13\linewidth}
    \centering
    \centerline{\scriptsize \textcolor{white}{g} \textcolor{white}{g}}
    \vspace{0.02cm}
    \includegraphics[width=\textwidth]{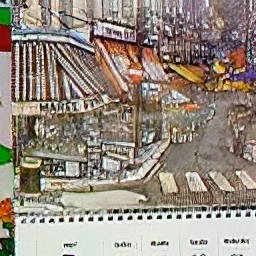}
    \vspace{-0.3cm}
    \centerline{\tiny 0.120$\mid${19.9}$\mid$0.688}\medskip
    \end{minipage}
    \hspace{0.15cm}
    \begin{minipage}[b]{0.13\linewidth}
    \centering
    \centerline{\scriptsize \textcolor{white}{g} \textcolor{white}{g}}
    \vspace{0.02cm}
    \includegraphics[width=\textwidth]{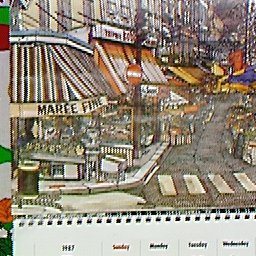}
    \vspace{-0.3cm}
    \centerline{\tiny\textcolor{white}{0.413$\mid$}}\medskip
    \end{minipage}    
    \vspace{.6em}       

        \vspace{-.5em}    
\caption{``LPIPS$|$PSNR$|$SSIM'' under each result. Single-codebook SVR was used. $x^{LQ}$ was computed by VVC with $qp\!=\!42$. The average $b_{LQ}\!=\!0.06$ and $b_c\!=\!0.035$. Our SVR-based LVC can largely improve reconstruction fidelity and perceptual quality.}
\label{fig:LVC-examples}
\end{figure*}

For video compression, we tested the single-codebook $\text{SVR}_s$. The LQ substitute $x^{LQ}$ was generated by the VVC standard \cite{vvc_std} using $qp\!=\!42$ over the original resolution. This is basically the lowest bpp configuration of VVC ($bpp\!=\!0.06$) with a reasonable reconstruction quality for $x^{LQ}$. Tab.~1 gives the rate-distortion performance, and Fig.~\ref{fig:LVC-examples} gives some restoration examples. Our ``$\text{SVR}_s$(LQ)" achieved much better perceptual quality with a 64.8\% improvement over LPIPS given only $0.035bpp$ increase. As expected, improvements over PSNR and SSIM are less significant as VVC is tailored to optimize such pixel-level distortions. With overall $bpp\!<\!0.1$, the SVR-based LVC can generate rich visually pleasing details compared to the overly smoothed results from VVC. On average, 47\% of codeword indices remain unchanged, verifying the effectiveness of using the similar  pipeline for both SVR-based LIC and LVC.

\setlength{\tabcolsep}{4pt}\vspace{-1em}
\begin{table}
\begin{center}
\caption{SVR-based LVC performance}\vspace{-1em}
\label{table:LVC-performance}
\begin{tabular}{ccccc}
\hline\noalign{\smallskip}
 & PSNR & SSIM & LPIPS & bpp\\
\noalign{\smallskip}
\hline
\noalign{\smallskip}
$\text{SVR}_s(LQ)$ & 28.15 & 0.812 & 0.109 & 0.095\\
VVC ($x^{LQ}$) & 28.09 & 0.806 & 0.310 & 0.06\\
\hline
\end{tabular}
\end{center}
\end{table}\vspace{-3.5em}
\setlength{\tabcolsep}{1.4pt}

\section{Conclusions}

We proposed a general SVR-based compression framework for both LIC and LVC. Based on the idea of guided image generation with conditional controls, our method drew fidelity cues as control signals from a low-bitrate LQ version of the original input to guide the reconstruction process. Compared with previous approaches that relied on SVR-based generation alone, the fidelity cues largely improved the reconstruction quality. By tuning the bitrate of the LQ input, we could trade off bitrate, reconstruction fidelity and perceptual quality. By transferring the difference of codeword indices between adjacent frames, a similar processing pipeline was used for both SVR-based LIC and SVR-based LVC. Experimental results showed improved performance over SOTA image and video compression methods. 


\clearpage
%
%
\bibliographystyle{splncs04}
\bibliography{egbib}
\end{document}